\documentclass[conference]{IEEEtran}

\usepackage{xcolor}
\usepackage{xspace}

\usepackage{hyperref} 

\usepackage{cite}
\usepackage{amsmath,amssymb,amsfonts}
\usepackage{algorithmic}
\usepackage{graphicx}
\usepackage{subcaption}
\usepackage{textcomp}

\usepackage{tablefootnote} 

\usepackage{stmaryrd} 
\usepackage{txfonts} 

\usepackage{pifont} 
\usepackage{mathtools} 

\usepackage{listings, listings-rust}
\let\labelindent\relax 
\usepackage{enumitem} 
\usepackage[breakable]{tcolorbox}

\newcommand{\ourfigwidth}{\linewidth}

\usepackage{booktabs}

\usepackage{multirow}
\usepackage{diagbox}

\usepackage{array} 

\usepackage{tikz}
\usepackage{pgfplots}
\usetikzlibrary{shapes.geometric, arrows.meta}

\ifdefined\confver
    \newcommand{\extendappendix}{Code examples illustrating each strategy are shown in the Appendix in the extended version of this paper~\cite{userstudy}.}
\else
    \newcommand{\extendappendix}{Code examples illustrating each strategy are shown in the Appendix~\ref{sec:apdxpattern}.}
\fi

\newcommand\ignore[1]{}
\definecolor{mygray}{HTML}{e3e6e8}
\lstdefinestyle{diff}{
  language=Rust,
  moredelim=[is][\color{red}]{--}{--},
  moredelim=[is][\color{darkgreen}]{++}{++},
}
\lstdefinestyle{highlightc}{
  language=C,
  moredelim=[is][\color{darkgreen!100}]{**}{**},
}
\definecolor{red}{rgb}{1,0,0}
\lstdefinestyle{customrust}{
  language=Rust,
  style=boxed,
  literate={*mut}{{{\textcolor{red}{*mut}}}}4
           {FILE}{{{\textcolor{red}{FILE}}}}4
           {c\_char}{{{\textcolor{red}{c\_char}}}}6
           {c\_int}{{{\textcolor{red}{c\_int}}}}5
           {c\_ulong}{{{\textcolor{red}{c\_ulong}}}}7
           {offset}{{{\textcolor{red}{offset}}}}6
            {*const}{{{\textcolor{red}{*const}}}}6
             {unsafe}{{{\textcolor{red}{unsafe}}}}6
              {libc}{{{\textcolor{red}{libc}}}}4
              {putchar}{{{\textcolor{red}{putchar}}}}7
              {puts}{{{\textcolor{red}{puts}}}}4
              {iswspace}{{{\textcolor{red}{iswspace}}}}8
              {*p2}{{{\textcolor{red}{*p2}}}}3
}
\lstdefinestyle{customc}{
  language=C,
  style=boxed,
  keywordstyle=\color{brown},
  commentstyle=\color{gray},
  identifierstyle=\color{violet}
}

\tcbuselibrary{listingsutf8} 
\tcbset{
    mylistingstyle/.style={
        colback=gray!10, 
        colframe=black!50, 
        boxrule=0.5mm, 
        arc=1mm, 
        boxsep=0mm, 
        left=0mm, right=0mm, top=0mm, bottom=0mm,
        outer arc=0mm,
        listing only, 
        listing options={%
            basicstyle=\ttfamily\scriptsize\bfseries,
            commentstyle=\color{gray},
            keywordstyle=\color{blue},
            backgroundcolor=\color{gray!1},
            lineskip=-0.4ex,
            language=C,
            style=highlightc,
        }
    }
}

\definecolor{mygray}{HTML}{e3e6e8}
\definecolor{darkgreen}{RGB}{0, 150, 0}

\definecolor{myblue}{RGB}{0, 0, 255}
\definecolor{myred}{RGB}{255, 0, 0}
\definecolor{mybrown}{RGB}{204, 102, 0}
\definecolor{mydarkgray}{RGB}{128, 128, 128}

\newcommand{\code}[1]{{\setlength\fboxsep{1pt}\colorbox{mygray}{\lstinline|#1|}}}

\newcommand\todo[1]{#1}

\definecolor{darkspringgreen}{rgb}{0.09, 0.45, 0.27}
\definecolor{darkdarkgreen}{rgb}{0.09, 0.25, 0.13}
\definecolor{deletecolor}{rgb}{0.66, 0, 0}
\definecolor{addcolor}{rgb}{0, 0.57, 0}

\usepackage[normalem]{ulem}
\newcommand\tododel[1]{} 
\newcommand\todoadd[1]{#1} 
\newcommand\tododelsec[1]{} 
\newcommand\todoaddsec[1]{#1} 
\newcommand\tododelcam[1]{} 
\newcommand\todoaddcam[1]{#1} 
\makeatletter
\newcommand\footnoteref[1]{\protected@xdef\@thefnmark{\ref{#1}}\@footnotemark}
\makeatother

    {\color{black}}

\newcommand\mywarning[1]{\PackageWarning{PAPER-WARNING}{#1}}

\newcommand{\laertes}{\textsc{Laertes}\xspace}
\newcommand{\crown}{\textsc{Crown}\xspace}
\newcommand{\florine}{\textsc{Flourine}\xspace}
\newcommand{\toolvert}{\textsc{Vert}\xspace}
\newcommand{\initialversions}{17}

\newcommand{\declall}{1261}
\newcommand{\declownfrac}{49.2}
\newcommand{\declbrwfrac}{50.8}

\newcommand{\refbrwmutfrac}{11.6}
\newcommand{\refbrwimmutfrac}{39.2}

\newcommand{\varsglobalfrac}{4.4}
\newcommand{\varsheapfrac}{31.4}
\newcommand{\varsstackfrac}{14.2}

\newcommand{\optioncompfrac}{63.1}

\newcommand{\optionallfrac}{8.3}

\newcommand{\dststringfrac}{62.5}
\newcommand{\dstbufferfrac}{16.4}

\newcommand{\dstpolyfrac}{0.8}
\newcommand{\dstallfrac}{79.7}

\newcommand{\safetyOwnTempstat}{95.6}
\newcommand{\safetyOwnTempdyn}{4.4}
\newcommand{\safetyOwnSpatstat}{1.6}
\newcommand{\safetyOwnSpatdyn}{98.4}

\newcommand{\safetyGlobalTempstat}{45.9}
\newcommand{\safetyGlobalTempdyn}{54.1}
\newcommand{\safetyGlobalSpatstat}{0.0}
\newcommand{\safetyGlobalSpatdyn}{100.0}

\newcommand{\safetyStackTempstat}{100.0}
\newcommand{\safetyStackTempdyn}{0.0}
\newcommand{\safetyStackSpatstat}{12.7}
\newcommand{\safetyStackSpatdyn}{87.3}

\newcommand{\safetyImrefTempstat}{100.0}
\newcommand{\safetyImrefTempdyn}{0.0}
\newcommand{\safetyImrefSpatstat}{5.5}
\newcommand{\safetyImrefSpatdyn}{94.5}

\newcommand{\safetyMutrefTempstat}{98.6}
\newcommand{\safetyMutrefTempdyn}{1.4}
\newcommand{\safetyMutrefSpatstat}{21.2}
\newcommand{\safetyMutrefSpatdyn}{78.8}

\newcommand{\safetyRefTempstat}{99.7}
\newcommand{\safetyRefTempdyn}{0.3}
\newcommand{\safetyRefSpatstat}{9.1}
\newcommand{\safetyRefSpatdyn}{90.9}

\newcommand{\safetyOptionTempstat}{100.0}
\newcommand{\safetyOptionTempdyn}{0.0}
\newcommand{\safetyOptionSpatstat}{10.8}
\newcommand{\safetyOptionSpatdyn}{89.2}

\newcommand{\safetyStringTempstat}{99.7}
\newcommand{\safetyStringTempdyn}{0.3}

\newcommand{\safetyBufferTempstat}{99.0}
\newcommand{\safetyBufferTempdyn}{1.0}
\newcommand{\safetyBufferSpatstat}{0.0}
\newcommand{\safetyBufferSpatdyn}{100.0}

\newcommand{\safetyPolyTempstat}{100.0}
\newcommand{\safetyPolyTempdyn}{0.0}
\newcommand{\safetyPolySpatstat}{0.0}
\newcommand{\safetyPolySpatdyn}{100.0}

\newcommand{\safetyDstTempstat}{99.6}
\newcommand{\safetyDstTempdyn}{0.4}

\newcommand{\cmpFuzzMinAllfail}{37}
\newcommand{\cmpFuzzMaxAllfail}{100}
\newcommand{\cmpFuzzAvgAllfail}{68}

\newcommand{\capiND}{63.3}

\newcommand{\liftingelision}{25}
\newcommand{\liftingsnapshot}{8}
\newcommand{\liftingcow}{5}

\newcommand{\safetyFlourineTestLine}{81.8}
\newcommand{\safetyFlourineTestPtrDecl}{88.2}
\newcommand{\safetyFlourineShocoLine}{96.3}
\newcommand{\safetyFlourineShocoPtrDecl}{100}
\newcommand{\safetyFlourineFmtLine}{73.1}
\newcommand{\safetyFlourineFmtPtrDecl}{100}
\newcommand{\safetyFlourineJoinLine}{74.3}
\newcommand{\safetyFlourineJoinPtrDecl}{100}
\newcommand{\safetyFlourineExprLine}{85.3}
\newcommand{\safetyFlourineExprPtrDecl}{100}
\newcommand{\safetyFlourineUrlparserLine}{100}
\newcommand{\safetyFlourineUrlparserPtrDecl}{100}

\newcommand{\safetyFlourinePrintfLine}{77.1}
\newcommand{\safetyFlourinePrintfPtrDecl}{100}
\newcommand{\safetyFlourineCsplitLine}{100}
\newcommand{\safetyFlourineCsplitPtrDecl}{100}
\newcommand{\safetyVertCsplitLine}{97.2}
\newcommand{\safetyVertCsplitPtrDecl}{100}
\newcommand{\safetyVertPrintfLine}{95.9}
\newcommand{\safetyVertPrintfPtrDecl}{100}
\newcommand{\safetyVertTestLine}{62.9}
\newcommand{\safetyVertTestPtrDecl}{84.6}
\newcommand{\safetyVertJoinLine}{94.3}
\newcommand{\safetyVertJoinPtrDecl}{100}
\newcommand{\safetyVertUrlparserLine}{84.1}
\newcommand{\safetyVertUrlparserPtrDecl}{78.6}
\newcommand{\safetyVertExprLine}{98.8}
\newcommand{\safetyVertExprPtrDecl}{93.1}
\newcommand{\safetyVertFmtLine}{60.1}
\newcommand{\safetyVertFmtPtrDecl}{100}
\newcommand{\safetyVertShocoLine}{99.4}
\newcommand{\safetyVertShocoPtrDecl}{100}

\newcommand{\safetyLaertesTestLine}{1.0}
\newcommand{\safetyLaertesTestPtrDecl}{9.5}
\newcommand{\safetyLaertesShocoLine}{1.2}
\newcommand{\safetyLaertesShocoPtrDecl}{0}
\newcommand{\safetyLaertesFmtLine}{1.2}
\newcommand{\safetyLaertesFmtPtrDecl}{3.0}
\newcommand{\safetyLaertesJoinLine}{0.9}
\newcommand{\safetyLaertesJoinPtrDecl}{14.3}
\newcommand{\safetyLaertesExprLine}{1.3}
\newcommand{\safetyLaertesExprPtrDecl}{5.0}
\newcommand{\safetyLaertesUrlparserLine}{0}
\newcommand{\safetyLaertesUrlparserPtrDecl}{0}
\newcommand{\safetyLaertesPrintfLine}{1.4}
\newcommand{\safetyLaertesPrintfPtrDecl}{3.0}
\newcommand{\safetyLaertesCsplitLine}{1.3}
\newcommand{\safetyLaertesCsplitPtrDecl}{3.2}
\newcommand{\safetyCrownTestLine}{1.6}
\newcommand{\safetyCrownTestPtrDecl}{9.5}
\newcommand{\safetyCrownShocoLine}{0.3}
\newcommand{\safetyCrownShocoPtrDecl}{0}
\newcommand{\safetyCrownFmtLine}{1.8}
\newcommand{\safetyCrownFmtPtrDecl}{3.0}
\newcommand{\safetyCrownJoinLine}{1.4}
\newcommand{\safetyCrownJoinPtrDecl}{2.4}
\newcommand{\safetyCrownExprLine}{1.8}
\newcommand{\safetyCrownExprPtrDecl}{2.5}
\newcommand{\safetyCrownUrlparserLine}{0.2}
\newcommand{\safetyCrownUrlparserPtrDecl}{1.3}
\newcommand{\safetyCrownPrintfLine}{2.1}
\newcommand{\safetyCrownPrintfPtrDecl}{3.0}
\newcommand{\safetyCrownCsplitLine}{1.8}
\newcommand{\safetyCrownCsplitPtrDecl}{3.2}

\newcommand{\unsafetyLaertesPtrDeclAvg}{95.3}
\newcommand{\safetyCrownPtrDeclAvg}{3.1}

\newcommand{\correctLaertesExpr}{95.3}

\newcommand{\correctLaertesTest}{100}
\newcommand{\correctCrownTest}{100}
\newcommand{\correctLaertesPrintf}{100}
\newcommand{\correctCrownPrintf}{100}
\newcommand{\correctLaertesJoin}{100}

\newcommand{\correctLaertesUrlparser}{100}
\newcommand{\correctCrownUrlparser}{100}

\newcommand{\correctCrownFmt}{100}
\newcommand{\correctLaertesShoco}{100}
\newcommand{\correctCrownShoco}{100}
\newcommand{\correctLaertesCsplit}{100}
\newcommand{\correctCrownCsplit}{100}

\newcommand{\realcodeTypeBmsCoverageRatio}{83}
\newcommand{\realcodeTypeStrRatiostr}{14}
\newcommand{\realcodeTypeStrRatioString}{18}
\newcommand{\realcodeTypeStrRatioOsStr}{4}
\newcommand{\realcodeTypeStrRatioOsString}{6}

\newcommand{\realcodeEquivPertLogical}{2.3}
\newcommand{\realcodeEquivPertErrmsg}{21.3}

\newcommand{\realcodeEquivPertLogicalAndErrmsg}{23.6}
\newcommand{\realcodeEquivPertFail}{50.6}

\begin{document}
\title{Translating C To Rust: Lessons from a User Study{\textsuperscript{\dag}}
}

\makeatletter
\newcommand{\linebreakand}{%
  \end{@IEEEauthorhalign}
  \hfill\mbox{}\par
  \mbox{}\hfill\begin{@IEEEauthorhalign}
}
\makeatother

\author{\IEEEauthorblockN{Ruishi Li\textsuperscript{*}}
\IEEEauthorblockA{\href{mailto:liruishi@u.nus.edu}{liruishi@u.nus.edu}\\
National University of Singapore\\
Singapore}
\and
\IEEEauthorblockN{Bo Wang\textsuperscript{*}}
\IEEEauthorblockA{\href{mailto:bo\_wang@u.nus.edu}{bo\_wang@u.nus.edu}\\
National University of Singapore\\
Singapore}
\and
\IEEEauthorblockN{Tianyu Li}
\IEEEauthorblockA{\href{mailto:tianyuli@u.nus.edu}{tianyuli@u.nus.edu}\\
National University of Singapore\\
Singapore}
\linebreakand
\IEEEauthorblockN{Prateek Saxena}
\IEEEauthorblockA{\href{mailto:prateeks@comp.nus.edu.sg}{prateeks@comp.nus.edu.sg}\\
National University of Singapore\\
Singapore}
\and
\IEEEauthorblockN{Ashish Kundu}
\IEEEauthorblockA{\href{mailto:ashkundu@cisco.com}{ashkundu@cisco.com}\\
Cisco Research\\
San Jose, CA, USA}
}

\IEEEoverridecommandlockouts
\makeatletter\def\@IEEEpubidpullup{6.5\baselineskip}\makeatother
\IEEEpubid{\parbox{\columnwidth}{
		Network and Distributed System Security (NDSS) Symposium 2025\\
		24-28 February 2025, San Diego, CA, USA\\
		ISBN 979-8-9894372-8-3\\
		https://dx.doi.org/10.14722/ndss.2025.241407\\
		www.ndss-symposium.org
}
\hspace{\columnsep}\makebox[\columnwidth]{}}

\maketitle

\begin{NoHyper} 
\renewcommand\thefootnote{\textsuperscript{*}}
\footnotetext{Contributed equally to this work.}
\renewcommand\thefootnote{\textsuperscript{\dag}}
\footnotetext{This is the extended version of the paper. The definitive Version of Record is to be published in NDSS 2025.}
\end{NoHyper}

\begin{abstract}
Rust aims to offer full memory safety for programs, a guarantee that untamed C programs do not enjoy. How difficult is it to translate existing C code to Rust? To get a complementary view from that of automatic C to Rust translators, we report on a user study asking humans to translate real-world C programs to Rust. Our  participants are able to produce safe Rust translations, whereas state-of-the-art automatic tools are not able to do so. Our analysis highlights that the high-level strategy taken by users departs significantly from those of automatic tools we study.
We also find that users often choose zero-cost (static) abstractions for temporal safety,
which addresses a predominant component of runtime costs in other full memory safety defenses. 
User-provided translations showcase a rich landscape of specialized strategies to translate the same C program in different ways to safe Rust, which future automatic translators can consider.
\end{abstract}

\section{Introduction}
\label{sec:intro}

Memory safety vulnerabilities in C programs are still a prominent category of CVEs reported in commodity software, and number in thousands each year~\cite{ossfuzz}. Several approaches to secure such unsafe code are being investigated.  

Full memory safety through runtime checks inserted by compiler instrumentation is achievable. However, it incurs high performance overhead ($\geq 50\%$) and is often not deployed in production~\cite{cets,softbound}. Instead, partial memory safety defenses which have overheads below $20\%$ have found deployment~\cite{cfi2014}. Emerging hardware features can reduce performance overheads of partial safety techniques, but even then, faulty programs that will often ungracefully abort in production systems have limited appeal.
Similarly, automatic patching techniques that can localize bugs and suggest 
fixes for them are being developed~\cite{faultloc,senx,gao2022programrepair,song2024provenfix}. But these approaches do not aim to rule out the existence of memory safety bugs altogether.

A different approach has been to {\em write secure code} which is free of memory safety bugs from the ground up. The idea is to have a stricter language and compiler that forces developers to rewrite unsafe code to use safe patterns everywhere. The advantage is that full memory safety can be achieved {\em mostly} statically, while a few runtime checks incur low overheads.

Designing safe C dialects with that goal has a long history. These dialects largely aim to retain low-level features of C for compatibility, making it difficult to move away from unsafe C patterns entirely while keeping overheads low~\cite{cyclone,necula2005ccured}. More recent designs aim for better compatibility by allowing mixed (statically) safe and unsafe C~\cite{checkedc}, but when temporal memory safety is desired in them, overheads can exceed $30\%$~\cite{3c}. 

Rust is an alternative mainstream language that offers low-level control over memory, while providing full memory safety. It has a growing ecosystem and increasing support from commodity OSes. It departs from the conventional approaches to safe C dialects in that it largely abandons the programming abstractions in C, such as the use of unchecked raw pointers and unsafe type casts. It is natural to ask: How difficult is it to translate existing C code to Rust then?  

In the last few years, automatic techniques to translate C code to Rust have started to emerge. There are two main approaches, one based on compiler-based analysis~\cite{emre2021translating,c2rust} and another based on large language models (LLMs)~\cite{eniser2024towards,yang2024vert}. But both approaches have had very limited effectiveness so far, even on small programs of about $100$ lines of code. For example, 
Emre et. al. report that only about $11\%$ of raw C pointers can be converted to safe Rust references through static analysis~\cite{emre2021translating}. Similarly, recent work on \florine~\cite{eniser2024towards} reports that
less than half of the small C programs they consider can be auto-translated to Rust using LLM-based repair.

It is unclear what strategies, if any, enable a successful search for C to Rust translations. In order to gain a complementary perspective, in this paper, we study how {\em human users approach the C to Rust translation task}. We conducted a user study in which we asked undergraduate students taking a course in computer security to translate a given set of real-world C programs to safe Rust. Our participants are familiar with C and memory safety issues, but have minimal prior exposure to Rust. To the best of our knowledge,
this is the first such analysis of a user study on C to Rust translation.

$33$ of our user study participants consented to their translated programs being analyzed and reported on. Most of the users succeed in providing reasonable translations of our C benchmark to Rust, whereas state-of-the-art automatic C to Rust translation tools are ineffective. 
\todoadd{Analyzing how users succeed is our main goal.} We highlight several high-level principles and strategies that are common across the user translations, which could be useful for future automatic translators. 

First, we find that all user-provided Rust translations {\em semantically lift} from the low-level abstractions used in the C code to re-express the logic in Rust, rather than trying to mimic the original program and data-flow structure too closely.
This approach helps break free of the low-level constraints present in the original C code, which would violate the strict rules in Rust if preserved. Rust enforces stricter rules on pointer aliasing than C. Users decomposed the object lifetimes in different ways that are {\em specialized to access patterns} used for the object to satisfy Rust rules. This shows that
there are multiple strategies to translate the same C program to safe Rust, with room for context-specific translation strategies.

Second, we find that resulting Rust translations often contain zero-cost statically-checked safety abstractions. Temporal memory safety,
which is a source of significant runtime overhead in many prior defenses~\cite{cets,3c}, is achieved mostly statically. The resulting Rust programs 
\todoadd{have comparable performance to C}, and rarely more than $20\%$ slower than the corresponding C code, even though our participants are not explicitly asked to optimize for performance. At the same time, known spatial and temporal vulnerabilities in our C benchmarks are eliminated in the translated Rust code. This suggests that C to Rust translation is worth it, as the trade-off between performance and security we observe in translations by non-expert Rust users is desirable.

Third, nearly all Rust translations have functional discrepancies compared with the corresponding C code. These discrepancies are easy to find with fuzzing and Rust translations of the same C program have correlated failures, i.e., they often fail the same tests. \tododel{Many}\todoadd{Some of these} discrepancies are expected since undefined behavior present in C programs are handled differently by different users. But, several others are logical errors, suggesting a ``last mile'' phenomenon at play: While users find it feasible to convert most C functionality to its Rust counterpart, making a $100\%$ correct translation is tedious.

\todoadd{Lastly, in order to examine the broader validity of our observations, we perform a post-hoc investigation of a mature open-source Rust project that mirrors the functionality of a set of popular C programs. Several findings from our user study, which involves non-experts, are also seen in the open-source project maintained by more experienced Rust developers.}

\noindent\textbf{Contribution.} We report on the lessons garnered from the first user study on translating real-world C programs to Rust by non-expert Rust users. Our work sheds light on strategies 
and policy decisions that users have to make, which are complementary to automatic mechanisms. We highlight the end gains in user-provided translations, such as the ubiquitous use of zero-cost temporal safety abstractions, \todoadd{elimination of unsafe code patterns, and the trade-offs with functional correctness.} \tododel{We hope our findings spur efforts in automatic C to Rust translation.} 

\section{Background}
\label{sec:background}

Rust allows programs to have fine-grained control over memory but with spatial and temporal memory safety.

\subsection{A Tour of Safe Rust}\label{sec:back_rust}
\label{sec:rust}

We briefly explain Rust design principles for achieving full memory safety with a running example shown in Figure~\ref{fig:introrust}.

\begin{figure}[t]
    \centering
    \includegraphics[trim=0cm 1.74cm 4.76cm 0cm, clip, width=\ourfigwidth]{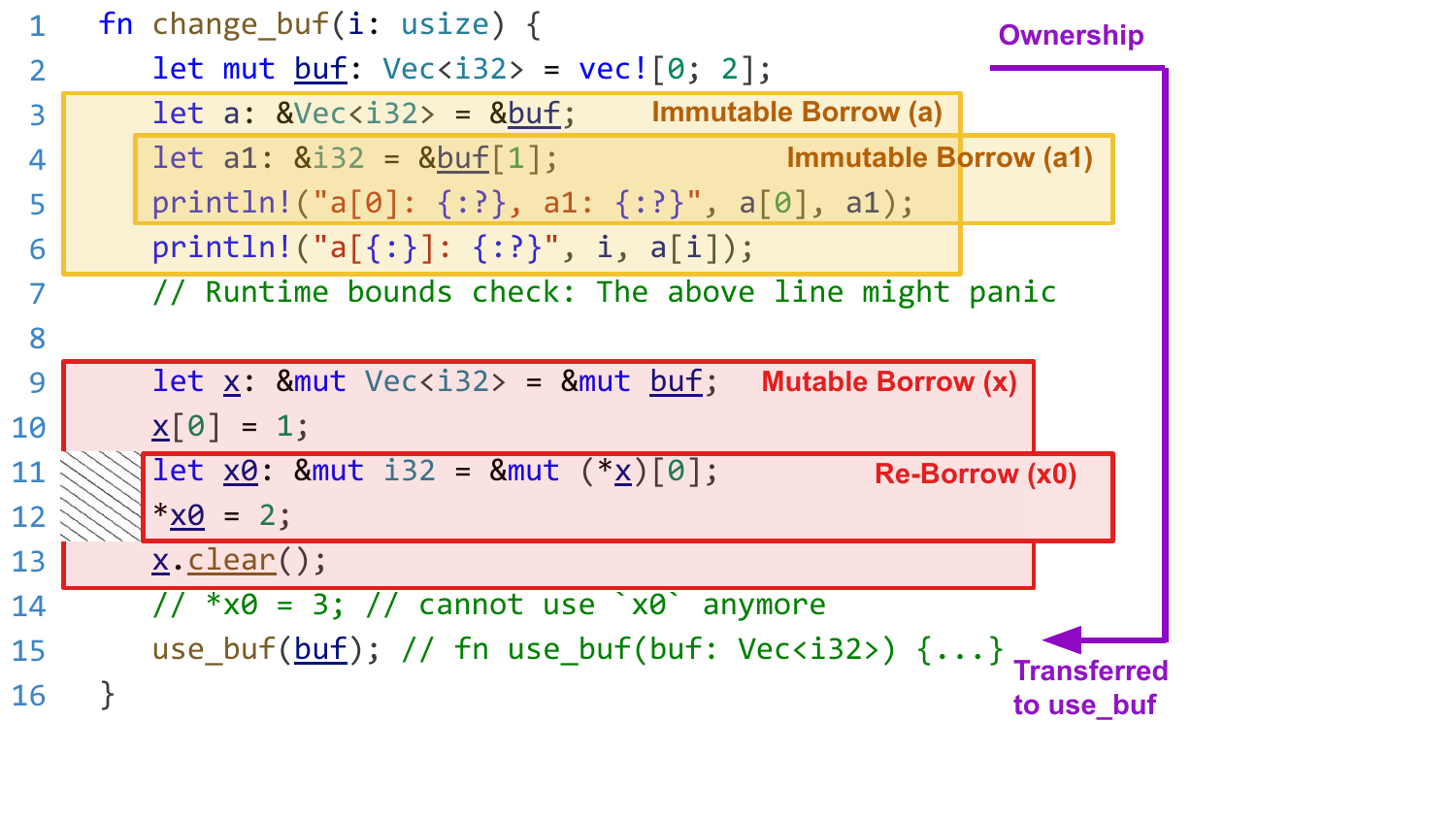}
    \caption{A Rust code example showing the concepts of ownership transfer, mut./immut. borrowing, and reference lifetime.}
    \label{fig:introrust}
\end{figure}

\noindent\textbf{Ownership.} 
Rust ensures that each data object has an exclusive {\em owner} at any given program point~\cite{ownershipilya}. 
When the variable owning an object goes out of scope, the compiler automatically deallocates the object. The single-owner principle ensures that the object is deallocated only once, avoiding double-free bugs prevalent in C programs. 
The compiler knows the scope of all objects in the program. Stack variables have statically determined block/function scope, and so do globals. 
Objects can be owners of other objects, creating a transitive chain
of ownership.
The chain starts with a stack/global variable, so the compiler can track ownership of all objects.
\mywarning{The actual definitions of ownership might be different. Check with a PL expert.}

Ownership can not be duplicated but only transferred between variables through assignment or parameter passing. The Rust compiler, therefore, tracks the unique owner of every object at every program point. 
For example, the \code{buf} variable in the example of Figure~\ref{fig:introrust} is on the stack and encapsulates a (smart) pointer to \code{Vec} object allocated on the heap. 
The object allocated on Line 2 in Figure~\ref{fig:introrust} has a unique owner \code{buf} from Line 2-14, after which the ownership transfers to the first parameter of \code{use_buf()}. The object will be deallocated when the new owner goes out of scope, i.e., at the end of the function scope of \code{use_buf()}.

\noindent\textbf{Borrowing.} Exclusive ownership, by itself, disallows all aliasing altogether, which is too restrictive. Rust relaxes this restriction by introducing a {\em borrowing} mechanism. It allows creating {\em references} (also called {\em borrows}) to temporarily access the data value without transferring the ownership. References \code{a}, \code{a1}, and \code{x} borrow from the owner \code{buf}, whereas \code{x0} {\em reborrows} from \code{x} in the example. This allows limited forms of aliasing.

\noindent\textbf{Lifetime.}
With borrowed references, it becomes essential to ensure that 
a reference never points to an object that has been deallocated.
The Rust compiler statically tracks the {\em lifetime} of variables: the region of code that the variable must be valid for. Rust enforces that the lifetime of all references is a strict subset of the lifetime of the owner statically. Since the object will be deallocated only when the lifetime of the owner expires, all live references point to valid objects, eliminating use-after-free vulnerabilities that arise in C programs. The example in Figure \ref{fig:introrust} creates multiple references to an array object. The variable \code{buf} gets ownership of the heap-allocated array on Line 2. There are four borrowed references created directly or indirectly from \code{buf} on lines 3, 4, 9, and 11. Their lifetimes are enforced to be smaller than that of \code{buf} and are highlighted. The object is automatically deallocated only when the lifetime of the owner \code{buf} ends, thereby ensuring that all references point to valid memory throughout their lifetime.

\noindent\textbf{Aliasing Xor Mutability (AXM).} 
Aliasing references can update the object storage, possibly moving them, and associated allocation metadata (e.g. size, internal pointers) can become inconsistent. When multiple references are pointing to the same object (e.g. a \texttt{vector} type) to insert/delete elements or change its capacity, the location of its internal buffer and size metadata might change. Consequently, this will invalidate other references to the same object, including iterators and references to its elements. Memory-unsafe languages like C leave careful management of aliased pointers \tododel{and their semantic metadata }to the programmer, which has been a source of mistakes. 
To avoid this, Rust proposes {\em Aliasing Xor Mutability (AXM)} principle.
First, references are divided into two types: mutable (\code{&mut}) and immutable (\code{&})\tododel{, prefixed with \code{&mut} and \code{&} declarations keywords respectively in Rust}. Mutable references can read and write the referent object, while immutable ones have read-only access.
Second, the Rust compiler enforces that either only one mutable reference or multiple (aliasing) immutable references are active at a program point. Mutable aliased references can not be active at the same program point. The AXM principle eliminates the need for careful pointer invalidation by the programmer and helps memory safety in concurrent code as well.

Specifically in the example of Figure~\ref{fig:introrust}, Lines 3-6 have multiple immutable references to \code{buf}, and Lines 9-13 have one mutable reference at each point, created using \code{&mut} type declaration. The compiler can statically infer the lifetime of \code{x} and other mutable references that may alias it, such as \code{x0}. It automatically splits their range of legal use such that there will be a single mutable reference to the object active in scope at each statement, i.e, \code{x} is permitted on Line $(9-10, 13)$ and \code{x0} on Lines 11-12.
The compiler would forbid using \code{x0}
on or after line 13 as it would create two mutable references to the vector elements, highlighted by commented code on Lines 14 for illustration. In the equivalent C code, dereferencing \code {x0} on Line 14 would be legal and lead to undefined behavior, since \code {x0} refers to a vector whose underlying elements were cleared (possibly deallocated) via \code{x} on Line 13. This can result in out-of-bound access due to a dangling dereference. Such memory errors are disallowed by the AXM principle in Rust. 

\noindent\textbf{Runtime AXM and Thread Safety.} 
The AXM principle can be too restrictive in certain circumstances. When multiple mutable references are necessary, Rust programs can resort to runtime borrow checks and reference counting for memory safety, through the use of data types such as \code{RefCell<T>} and \code{Rc<T>} \todoadd{for single-thread synchronization~\cite{threadlocalstorage} and \code{Mutex<T>} and \code{Arc<T>} for multi-thread synchronization.}
These generic types are implemented in the Rust standard library with internal use of unsafe Rust and have been proven sound~\cite{rustbelt}. Mutable pointers are particularly common in multi-threaded code, and in fact, concurrent C code is notoriously difficult to analyze for memory safety bugs.
The Rust compiler enforces that concurrent access or data copying across threads will never result in memory safety violations. 
Thus, when Rust allows relaxations to the AXM principle, runtime checks are used for memory safety. We will refer to all such references that carry dynamic checks as {\em dynamic references} to avoid confusion with regular Rust references that obey the AXM principle.

\noindent\textbf{Spatial Bounds Checking.}
The Rust compiler aims to ensure that the accesses by owners or references to the referent are within spatial bounds.
For several types of objects, the compiler can do so statically, but when it cannot, it adds runtime bounds checks. 
In line 6 of Figure~\ref{fig:introrust}, \code{a[i]} operation is checked to see if \code{i} is within the boundary of \code{buf}. If passing 3 to \code{i}, the Rust code would abort the execution of the program. 

\noindent\textbf{Type Safety.}
Rust is a type-safe language and supports safe type casting among data types using APIs provided in the standard library. All type casts guarantee that the converted values are within the legal range of the target type. Safe Rust disallows direct reinterpretation of memory to an unsafe type.
For example, the conversion between pointers (\code{void\*}), non-primitives (\code{struct}, \code{union}, etc.) and primitives (\code{bool}, \code{enum}, etc.) can cause memory errors in C, but \todoadd{safe Rust} forbids them.

\subsection{Existing Automatic C to Rust Translation}

There have been $2$ main approaches to automatic C-to-Rust translation:
One based on compiler or static analysis, while the other using large language models (LLMs) for code synthesis.
\noindent\textbf{Compiler-based Approach.} Most of the work in this direction decomposes the code translation problem into two stages: (1) C to unsafe Rust, and (2) unsafe Rust to safe Rust~\cite{emre2021translating,emre2023aliasing,zhang2023ownership}. 
\todoadd{A mature tool called \texttt{c2rust} performs a robust line-by-line syntactic translation from C to corresponding \code{unsafe} Rust code blocks~\cite{c2rust}.}
The use of \code{unsafe\{...\}} keyword allows bypassing the Rust safety checks mentioned above in the enclosed code block. The resulting Rust translation can thus rely on raw pointers (\code{mut*}), C-compatible data types, and foreign function calls to C libraries (\code{extern "C" fn}). 
The second stage of the translation is where the bulk of the challenges lie. 
The goal of prior automatic tools is to refactor the unsafe Rust code into safe Rust and reduce unsafe code. 
\textsc{Laertes} proposed a trial-and-error approach leveraging Rust compiler feedback to lift a certain subset of raw pointers into safer Rust references~\cite{emre2021translating}. 
However, a later work reports that the majority of raw pointers (79\% in their benchmarks) cannot be translated to safe references via such techniques~\cite{emre2023aliasing}. 
\textsc{Crown} proposed improvements using ownership analysis based on constraint generation and SAT-solving, for a subset of pointer types such as \code{Box<T>}~\cite{zhang2023ownership}. We present an evaluation of these state-of-the-art tools later and find that a majority of C pointers are not lifted into safe Rust abstractions by present tools.

\noindent\textbf{LLM-based Approach.}
Another line of work uses LLMs to translate C programs to Rust~\cite{eniser2024towards,yang2024vert}. 
The LLM-guided approach tends to produce code that is much more readable and idiomatic. Modern LLMs can suggest safe Rust data types, APIs, and coding conventions to use. However,
LLM-based translation is often difficult to control as LLMs can make semantic mistakes~\cite{pan2024lost}.
One of the most recent tools called \textsc{Flourine}, reports that with current LLMs (GPT-4, claude3, and so on), less than 20\% for C programs longer than 150 lines of code can be satisfactorily translated to Rust~\cite{eniser2024towards}.
We evaluate \textsc{Flourine} as well, and we have similar findings.

In summary, existing automatic techniques are insufficient to translate even small C programs of about $200$ LoC 
to Rust. Our observations from the user study hope to offer a holistic perspective on how users get past the inherent challenges. 

\section{User Study}
\label{sec:userstudy}

Prior literature has taken a \emph{bottom-up} approach to the problem of C to Rust translation, aiming to show how automatic tools can address particular sub-problems encountered in translation. Our study offers a complementary perspective on how human users approach the same task. This gives us a {\em top-down} view of the problem: We can see what common strategies do users use and what challenges remain thereafter.

Our participants are undergraduate students enrolled in a course on computer security. All participants are familiar with memory safety errors in C/C++ programs and were given an introduction to Rust\footnote{The content described in Section~\ref{sec:rust} with example exercises constitutes the introduction given, after the participants are familiar with C memory errors.}. 
The task, which is part of a graded course project, is to translate $8$ real-world C programs we collected into safe Rust. Each participant is randomly assigned one out of $8$ such C programs and is asked to provide a translated program in safe Rust within 20 days\footnote{The participants conducted their work
in April 2024.}. The C programs are taken from GitHub and are small due to the 20-day time limit, with lengths varying from 322 to 536 LoC. The participants were permitted to consult the web freely and use any existing tools, for example, \texttt{c2rust} and LLMs. 

\noindent\textbf{Ethical Concerns.} 
\todoaddsec{This study has been granted an IRB exemption from the NUS School of Computing DERC (Department Ethics Review Committee).}
We followed the procedure advised by the IRB to protect the privacy of the participants and avoid bias. $73$ undergraduate students who undertook the study were initially invited to give consent towards the use of their submissions. $33$ participants gave their consent and only their data is included in this study. The participant data was anonymized and kept confidentially on our research infrastructure, without being hosted on third-party cloud services. Aggregate statistics are reported here as far as possible, and wherever code snippets are shown for illustration of a concept, they are constructed synthetically by retaining the high-level patterns seen in user submissions (not replicated). Analysis of the participants' data was conducted only after grades were finalized to avoid any influence resulting from the study on the grades.

\begin{table}[ht]
    \caption{C Benchmarks for the User Study. }
    \label{tab:programs}
    \centering
    \begin{tabular}{|l|l|l|l|}
        \hline
        \textbf{Prog. Name} & \textbf{LoC} & \textbf{Description} \\
        \hline
        \href{https://github.com/DiegoMagdaleno/BSDCoreUtils/tree/d2b28e08bd02da5076a876608d6431638f929849/src/csplit}{\texttt{csplit (bsd\footnoteref{note1}})} & 322 & Split files based on patterns  \\
        \href{https://github.com/DiegoMagdaleno/BSDCoreUtils/tree/d2b28e08bd02da5076a876608d6431638f929849/src/expr}{\texttt{expr (bsd)}} & 451 & Evaluate expressions  \\
        \href{https://github.com/DiegoMagdaleno/BSDCoreUtils/tree/d2b28e08bd02da5076a876608d6431638f929849/src/fmt}{\texttt{fmt (bsd)}} & 415 & Format text files  \\
        \href{https://github.com/DiegoMagdaleno/BSDCoreUtils/tree/d2b28e08bd02da5076a876608d6431638f929849/src/join}{\texttt{join (bsd)}} & 472 & Join lines of two files  \\
        \href{https://github.com/DiegoMagdaleno/BSDCoreUtils/tree/d2b28e08bd02da5076a876608d6431638f929849/src/printf}{\texttt{printf (bsd)}} & 375 & Print formatted text  \\
        \href{https://github.com/DiegoMagdaleno/BSDCoreUtils/tree/d2b28e08bd02da5076a876608d6431638f929849/src/test}{\texttt{test (bsd)}} & 536 & \ignore{Check and compare file attributes} Check file attributes and values  \\
        \href{https://github.com/Ed-von-Schleck/shoco/tree/4dee0fc850cdec2bdb911093fe0a6a56e3623b71}{\texttt{shoco}} & 388 & \ignore{Compress text using shoco algorithm} String Data Compression  \\
        \href{https://github.com/jwerle/url.h/tree/a65623ad107be19ca4efb5a36379f3440eb48091}{\texttt{urlparser}} & 437 & Parse URLs  \\
        \hline
    \end{tabular}
\end{table}

\noindent\textbf{Benchmarks.}
Table \ref{tab:programs} shows the C programs we collected for the user study. 
These programs are representative of C programs that implement lower-level functionality and self-manage memory where memory safety errors arise. 
Such functionalities are often implemented in C and used by higher-level software systems.
We also considered the translation difficulty and chose programs between 300 and 600 LoC, considering the time and effort of the participants.
Thus, we chose $6$ C programs\footnote{\label{note1}Obtained from the \texttt{BSDCoreUtils}~\cite{bsdcoreutil} code repository (version \href{https://github.com/DiegoMagdaleno/BSDCoreUtils/tree/d2b28e08bd02da5076a876608d6431638f929849}{d2b28e0}).} from the \texttt{BSDCoreUtils}~\cite{bsdcoreutil} collection of system utilities and $2$ libraries, i.e., \texttt{shoco} and \texttt{urlparser}.
All of the chosen programs (Table \ref{tab:programs}) are self-contained, requiring only the C standard library as a dependency.
Four programs---\texttt{csplit}, \texttt{fmt}, \texttt{join}, and \texttt{test}---perform file processing.
The \texttt{expr} and \texttt{printf} are pure computation utilities for strings and numbers.
The \texttt{shoco} library is for data compression, and the \texttt{urlparser} library parses URL strings. 

\noindent\textbf{Task Requirements.}
Each individual participant is tasked to translate the assigned C program into safe Rust, satisfying two requirements, i.e., \emph{safety} and  \emph{functional correctness}.

\begin{enumerate}
    \item \textbf{Safety.} The translated program must be written in safe Rust only. The use of keyword \code{unsafe} is strictly forbidden. For dependencies, only the Rust standard library should be used by default. When that is insufficient to implement certain functionalities, additional third-party dependencies can be used if they are well-maintained.
    \item \textbf{Functional correctness.} The translated Rust program should have the equivalent external behavior to the C source program. 
    Since most of the chosen C programs do not come with high-quality unit tests by default, we asked the participants to write their own tests with line coverage aiming for at least 85\% on the original C program to test correctness. If the source program and the translation behave the same (i.e., output, return code, effects on the file system, etc.) on those tests, we say that the translated program is \todoadd{\emph{correct} if it passes all test cases created by users}.  \todoadd{Besides the final version of their translated program,} we also asked the participants to \todoadd{ submit an \emph{initial version} of the translation that could compile before they developed tests}, so we could analyze the changes.
\end{enumerate}

\noindent\textbf{Collected Translations.}
The 33 participants provided 31 {\em final} translations that can compile and achieved 70\% to 98\% line coverage on the C program under their respective tests.
26 of the 31 translations pass tests that reach $85\%$ coverage and nearly all of them\footnote{Except for 2 translations with coverage less than 80\%.} have coverage above $80\%$.
All of the 31 \todoadd{final} translations are written in pure \emph{safe} Rust, without the use of \code{unsafe}. They form the main target of our analysis. $\initialversions$ participants also submitted their compilable {\em initial} versions.

\noindent\textbf{Our Goal.} We analyze these $31$ \todoadd{final} translations to gain insights into the following research questions:\\
\noindent\textbf{RQ1.} Is there a set of common strategies that users used for successful translation?\\
\noindent\textbf{RQ2.} How is the \todoaddcam{security-performance} trade-off in the Rust translation?\\
\noindent\textbf{RQ3.} What are the common errors and correctness gaps in translated Rust programs?\\
\noindent\textbf{RQ4.} How do the state-of-the-art automatic C to Rust translation tools perform on the same task?

\section{High-Level Approach Taken by Users}
\label{sec:trans}

\begin{figure}[t]
    \centering\footnotesize
\begin{minipage}{\linewidth}
\begin{lstlisting}[language=C, style=customc]
static void center_stream(FILE *stream, const char *name)
{
    char *p1, *p2; // aliased pointers
    size_t len; int w1, w2; wchar_t wc;
    // ...
    while ((p1 = get_line(stream)) != NULL) {
        len = 0;
        for (p2 = p1; *p2 != '\0'; p2 += w2) {
            if (*p2 == '\t')
                *p2 = ' ';
            // ... skipped 
            if (len == 0 && iswspace(wc))  p1 += w2;
            else  len += w1;
        }
        while (l < goal_length) {
            putchar(' ');
            len += 2;
        }
        puts(p1);
    } ...
}
\end{lstlisting}
\end{minipage}
\texttt{Example C Program (above) $\rightarrow$ \texttt{c2rust} Translation (below)}\\
\begin{minipage}{\linewidth}
\begin{lstlisting}[language=Rust, style=customrust]
unsafe extern "C" fn center_stream(
  mut stream: *mut FILE, mut name: *const libc::c_char) 
{
    let mut p1: *mut libc::c_char = 0 as *mut libc::c_char;
    let mut p2: *mut libc::c_char = 0 as *mut libc::c_char;
    let mut wc: wchar_t = 0; // ...
    loop {
        p1 = get_line(stream);
        if p1.is_null() { break; }
        len = 0 as libc::c_int as size_t;
        p2 = p1;
        while *p2 as libc::c_int != '\0' as i32 {
            if *p2 as libc::c_int == '\t' as i32 {
                *p2 = ' ' as i32 as libc::c_char; }
            // ...
            if len == 0 ... && iswspace(wc as wint_t) != 0 {
                p1 = p1.offset(w2 as isize);
            } else {  len = ...;  }
            p2 = p2.offset(w2 as isize);
        }
        while len < goal_length {
            putchar(' ' as i32);
            len = (len as libc::c_ulong).wrapping_add(2) ...;
        }
        puts(p1);
    } ...
\end{lstlisting}
\end{minipage}
    \caption{(Top) An example of C program with aliasing pointers, i.e., multiple pointers pointing to the same region of memory. \todo{(Bottom) A line-by-line translation by \texttt{c2rust}. Uses of unsafe raw pointers and \code{libc} APIs/types are highlighted in red.}}
    \label{fig:exp_lifting_c}
\end{figure}

Fig.~\ref{fig:exp_lifting_c} shows an example of C code from the \texttt{fmt} benchmark and the translation produced by the \texttt{c2rust} tool.
The latter is created by preserving the structure of the original code, line-by-line and variable-to-variable, but contains unsafe Rust code blocks. The main challenge is to refactor the unsafe code to safe Rust. A straightforward removal of the keyword \code{unsafe} does not work as there are two immediate challenges:
\begin{itemize}
    \item[a)] the unsafe Rust code, much like the original C code, uses low-level raw pointers and C library calls that are disallowed in safe Rust; and
    \item[b)] even if there exists a way to replace all pointers with references, the resulting code will violate safe Rust rules.
\end{itemize}

In fact, for our shown example, there is no translation that keeps the original control-flow structure as-is and maps all the original C pointers to Rust references. This is because preserving the original lifetime and read/write semantics of the C variables in the Rust code will always violate the AXM principle, since \code{p2} needs to be a mutable reference and remain alive from Lines 8-14, and that interferes with the use of \code{p1}. 

Despite this apparent challenge, there are ways out of the quandary, and our participants translated such programs correctly. We explain what strategies they used and how often.

\subsection{Semantic Data Type Lifting}
\label{sec:lifting}

\begin{figure}[t]
    \centering
\begin{minipage}{\linewidth}
\begin{lstlisting}[language=Rust, style=boxed]
fn center_stream<R: BufRead>(mut stream: R, _name: &str, config: &Config) {
    let mut p1 = String::new();
    while let Ok(bytes_read) = get_line(stream, &mut p1) {
        if bytes_read == 0 { break; }
        let len: usize = p1.trim().chars().map(|c| if c == '\t' { ' ' } else { c }).map(char::len_utf8).sum();
        let padding = (config.goal_length - len) / 2; // Calculate padding to center the line
        for _ in 0..padding { print!(" "); }
        println!("{}", p1.trim());
    } ...
\end{lstlisting}
\end{minipage}
    \caption{A \tododel{Valid}\todoadd{safe} Rust translation of the C program (Version A) 
    }
    \label{fig:exp_lifting_rs1}
\end{figure}

\begin{table}[t]
    \caption{Dissimilarities between source C programs and Rust translations. Numbers are averaged over all  translations.}
    \label{tab:dissim}
    \centering
    \begin{tabular}{|l|c|c|c|c|}
\hline
\textbf{Benchmark} & \textbf{\#user-} & \textbf{\#function} & \textbf{\#signature} & \textbf{\#pointer decl.} \\
                   &  \textbf{translations}              & \textbf{(kept/rm./add.)} & \textbf{(sim./dissim.)} & \textbf{(kept/rm.)} \\
\hline
\texttt{csplit} & 5 & 7 / 2 / 3 & 5 / 2 &  54\% / 46\% \\
\hline
\texttt{expr} & 2 & 8 / 10 / 8 & 4 / 4 & 61\% / 39\% \\
\hline
\texttt{fmt} & 4 & 10 / 2 / 2 & 6 / 5 & 82\% / 18\% \\
\hline
\texttt{join} & 2 & 6 / 6 / 11 & 4 / 2 & 70\% / 30\% \\
\hline
\texttt{printf} & 3 & 11 / 2 / 4 & 5 / 6 & 71\% / 29\% \\
\hline
\texttt{test} & 4 & 16 / 2 / 3 & 10 / 6 & 89\% / 11\% \\
\hline
\texttt{shoco} & 6 & 6 / 0 / 2 & 2 / 5 & 72\% / 28\% \\
\hline
\texttt{urlparser} & 5 & 18 / 3 / 3 & 3 / 15 & 79\% / 21\% \\
\hline
    \end{tabular}
\end{table}

\begin{figure}[t!]
    \centering
    \includegraphics[trim=2.2cm 2.5cm 3.8cm 1.2cm, clip, width=\linewidth]{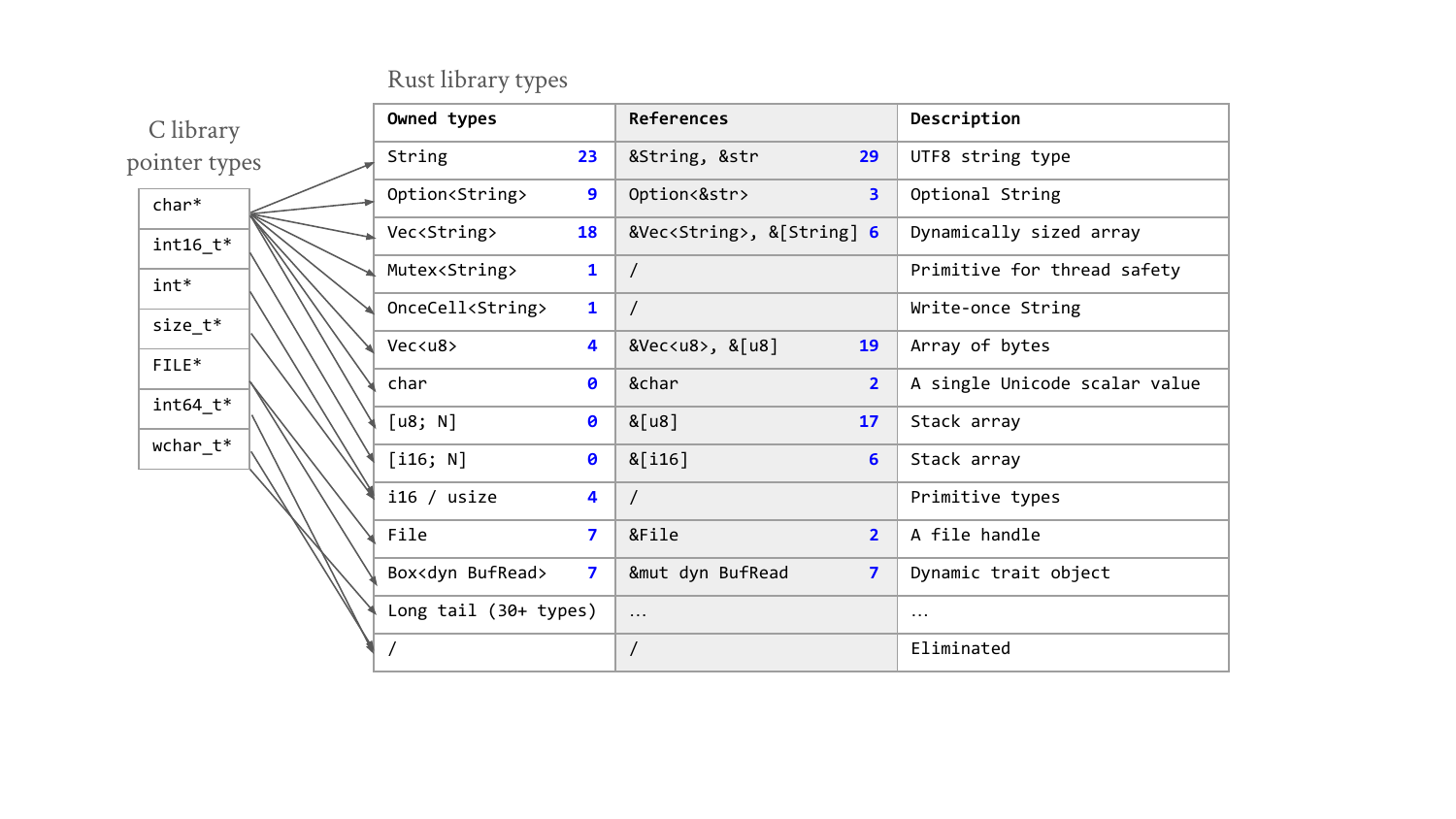}
    \caption{Raw C pointers to Rust data types lifted in translations \todoadd{and the number of programs using each Rust type (in \textcolor{blue}{blue})}.} 
    \label{fig:typemap}
\end{figure}

Several users recognize that the low-level \code{char*} pointers used in the example C code are semantically operating on a higher-level data type and they {\em lift} the object's type to a relevant Rust abstract data type. Figure~\ref{fig:exp_lifting_rs1} shows one such successful translation that lifts the \code{char*} buffer in C to \code{String} type. 
The C code with library calls is replaced with invocations to the Rust \code{String} methods. For example, the C code in Lines 6-20, which is responsible for counting the whitespace
characters and center-justifying the string, is implemented
with safe Rust \code{String} methods in the translation.
The result of such type lifting is that the Rust code is less similar to the original C code in the control-flow structure and variables used. \todoadd{Table}~\ref{tab:dissim} summarizes the amount of dissimilarity we observe in $31$ final translations and their corresponding C code.

Multiple choices for the Rust data types exist here. For example, \code{char*} may be lifted to Rust \code{String}, \code{Vec<u8>}, \code{Box<[u8]>}, and so on. 
\todo{Figure~\ref{fig:typemap} shows the top Rust data types} that our participants lifted \todo{from C pointers} across $8$ benchmarks in their final $31$ translations. Note that there are subtle differences between the original C type and the corresponding lifted data type that the user chose. In Version A, the code uses a \code{String} type, which does not support strings with invalid UTF-8 characters (unlike the original C code). The corresponding Rust translation elides that particular behavior present in the original C code. Section~\ref{sec:testdiff} quantitatively analyzes such semantic discrepancies in more detail.

\begin{figure}[t!]
    \centering
    \includegraphics[trim=2.5cm 2.6cm 8cm 4.8cm, clip, width=\ourfigwidth]{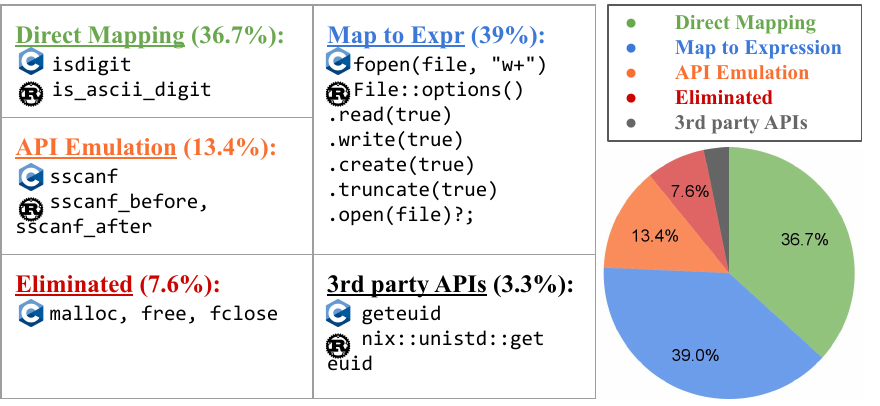}
    \caption{Breakdown of C library API translations with examples.}
    \label{fig:fig_api_mapping}
\end{figure}

A different sub-challenge with type lifting is how to translate operations on the original data type, which is often implemented with 
standard C library API, to corresponding Rust data type methods. 
\tododel{Fig.~\ref{fig:fig_api_mapping} shows the fraction of the C library operations for which our users found similar methods on standard Rust data types, across all $31$ \todoadd{final} translations. \todo{\capiND\% of} C library functions had no direct mapping to an API of Rust data types that the participants chose. Participants resorted to the following four strategies:}
\todoadd{Foreign function interface (or FFI) calls to C libraries are disallowed in safe Rust. User translations used the strategies below to translate C library API calls:
\begin{itemize}
    \item[(a)] \emph{Direct Mapping to Safe Rust APIs.} $36.7\%$ of the API calls are translated to single safe Rust method calls.
    \item[(b)] \emph{Translating to expressions.} $39\%$ of the API calls are translated into expressions that make use of multiple Rust methods and operators to achieve similar effects.
    \item[(c)] \emph{API Emulation.} $13.4\%$ of the API calls are translated into calls to user-implemented functions that emulate them.
    \item[(d)] \emph{Elimination.} $7.6\%$ of the calls for manual memory and resource management are eliminated since Rust data types can automatically achieve similar effects.
    \item[(e)] \emph{Third-party Libraries.} $3.3\%$ of the calls are translated to API calls provided by third-party libraries (crates).
\end{itemize}}
The breakdown of these choices with examples is summarized in Figure~\ref{fig:fig_api_mapping}. 
As a result of the above challenges, we observed that
the initial guess of the lifted data type chosen by a user was often not optimal. Of the $\initialversions$ participants who shared their initial version of Rust code, \todo{10} refined or changed their initial types chosen to other ones in the final version.

\begin{tcolorbox}[breakable,width=\linewidth,boxrule=1pt,top=0pt, bottom=0pt, left=1pt,right=1pt]
User-provided translations break away from the low-level structure of the original C code by semantically mapping C data types and APIs
to similar ones in Rust. They often refine their initial guess of such C to Rust mappings.
\end{tcolorbox}

\subsection{Dealing with Aliasing}
\label{sec:aliasing}

\begin{figure}[t]
    \centering
\begin{minipage}{\linewidth}
\begin{lstlisting}[language=Rust, style=customrust]
fn center_stream<R: BufRead>(mut stream: R, name: &str) {
    // ...
    while let p1: Vec<char> = get_line(stream) {
        for c in p1 {  // implicit mutable borrow by the loop
            if c == '\t' {
                c = ' ';
            } ...
            if (l == 0 && isWhiteSpace(wc))
                p1.drain(..wcl); // AXM violation (FAIL!)
            // ...
        } ...
        println!("{:?}", p1);
    } ...
\end{lstlisting}
\end{minipage}
    \caption{A Line-by-line Rust translation that fails to compile.}
    \label{fig:exp_lifting_rs01}
\end{figure}

Fig.~\ref{fig:exp_lifting_rs01} shows another example of translation wherein, after data types have been lifted, the Rust compiler checks are violated. 
This example illustrates a separate challenge: Users have to decide how each one of the original C pointers should be mapped to Rust references while satisfying Rust borrowing rules in the presence of aliases (borrows). 
Rust provides the option to use reference-counted dynamic references, which have accompanying runtime costs, but we see that users did not use them for heap and stack data references. Instead, users found $2$ strategies, specialized to the access patterns used by the C program, that satisfy static checks of the Rust compiler.

\noindent\textbf{Strategy (a): Elision}. When objects are lifted from C to Rust, many of the original C pointers do not need to be mapped to Rust references and can be elided in the translated code. For example,  code in Version A (Figure~\ref{fig:exp_lifting_rs1}) elides the C pointer \code{p2} because its functionality is encapsulated by Rust \code{String} methods. Version A \tododel{compiles and }satisfies the Rust compiler checks. 

\noindent\todoadd{\textbf{Strategy (b): Cloning.}
The translation shown in Version B (Figure~\ref{fig:exp_lifting_rs2}) embodies a different strategy. It satisfies the Rust borrow checking rules by separating the read and write accesses of the original object into two separate objects.
In Figure~\ref{fig:exp_lifting_rs01}, the \code{p1} reference and the iterator are simultaneously trying to modify the object, hence violating the AXM principle.
In Version B, however, a new copy of the original \code{String} object is made.
The \code{p1} reference can thus remain immutable and refer to the original object, while all writes are made to the copy 
by the mutable reference \code{ans}. Later on, only the writable copy of the object is used. Such a rewrite mechanism satisfies the borrow rules in safe Rust and compiles successfully.
}

\todoadd{
Two pointers to a read-only object can both access the object without any restriction in C. But in Rust, if one reference is (re)borrowed from the other, it should have a statically determined shorter lifetime than that of the other, which is more restrictive than C. The cloning strategy is thus also useful when translating C code with multiple immutable pointers. It disentangles the lifetime of two immutable pointers referring to an object by making a copy of it with the same value.}

\tododel{\noindent\textbf{Strategy (b): Copy-on-Write}. 
The translation Version B in Fig.~\ref{fig:exp_lifting_rs2} embodies a different strategy. It satisfies the Rust borrow checking rules by separating
the read and write accesses of the original object into two separate objects---similar to the idea employed in the concept of ''copy-on-write" (CoW).
In Fig~\ref{fig:exp_lifting_rs01}, the \code{p1} reference and the iterator are simultaneously trying to mutate the object, hence violating the AXM principle.
In Version B, however, a new copy of the original \code{String} object is made.
The \code{p1} reference can thus remain immutable and refer to the original object, while all writes are made to the copy 
by the mutable reference \code{ans}. Later, only the copy object is used. Such a rewrite mechanism
satisfies the borrow rules in safe Rust and compiles
successfully.
}

\tododel{
The CoW semantics can be used when the two copies can eventually be merged without violating any lifetime or semantic constraints. The original C code is not concurrent, and it is safe to create a temporary copy. This is why the translated code uses the \code{ans}, instead of \code{p1}, after line 18.
}

\begin{figure}[t]
    \centering
\begin{minipage}{\linewidth}
\begin{lstlisting}[language=Rust, style=boxed]
fn center_stream<T: BufRead>(&mut self, stream: T, name: &str) {
    for line in stream.lines() {
        let p1 = match line {
            Ok(line) => line,  Err(e) => { ... }
        };
        let mut ans = String::new();
        let mut len = 0;
        for c in p1.chars() {
            // ... if c is a space, skip
            if c == '\t' {
                ans.push(' ');
            } else {
                ans.push(c);
            }
            len += c.width().unwrap_or(1);
        }
        println!("{:>width$}", ans, width = ...);
    ...
\end{lstlisting}
\end{minipage}
    \caption{Possible translations of the C program (Version B) 
    }
    \label{fig:exp_lifting_rs2}
\end{figure}

\tododel{
\noindent\textbf{Strategy (c): Cloning}.
A related strategy used to satisfy borrowing rules is to create a \emph{clone}, a copy of an object for read accesses only. In Rust, if we transfer ownership, the original variable can no longer access the object since ownership cannot be duplicated.
When multiple places need to access the same data with differing lifetime constraints, and all accesses are read-only, cloning the object is a solution. We have elided an illustrative example here since it is similar to the CoW case. 
}

\noindent\textbf{How often are the above strategies used?} 
\emph{Elision} is the most frequently used specialization strategy when translating code fragments involving aliasing references. We find that it is used in \todo{\liftingelision} final translations.  \emph{Cloning} is used in \tododel{\todo{\liftingcow} and \todo{\liftingsnapshot}}\todoadd{13} translations. 

\begin{tcolorbox}[breakable,width=\linewidth,boxrule=1pt,top=0pt, bottom=0pt, left=1pt,right=1pt]
Rust translations provided by users choose specialized strategies to satisfy static Rust safety rules, rather than resort to dynamic references (ref-counted), to handle aliasing.
\end{tcolorbox}

\section{Security and Performance of User Translations}

One of the most important motivations for translating C to Rust is to guarantee full memory safety without sacrificing performance.
Some Rust safety abstractions are completely static, thereby having zero runtime costs, while others employ runtime checks. We analyze the usage of Rust abstractions in the translations obtained in our study to understand how often zero-cost safety abstractions are used. We then measure end-to-end performance of the Rust translations.
\todoadd{We also highlight prominent examples of code patterns known to be dangerous in C. These are forbidden in safe Rust, and we explain how they were translated to safe Rust code by our participants.}

\subsection{Breakdown of Safe Abstractions Used}
\label{sec:safety}

Rust offers smart references (pointers) and safe data type abstractions which can replace raw pointers in C. We analyze (1) how often the Rust translations use those data type abstractions that encapsulate pointers, and (2) whether the memory safety properties on those types are enforced at compile time (statically) or run time (dynamically).

\begin{table}[t]
    \centering
    \caption{
    Temporal and spatial memory safety of data references used in Rust translations. 
    }
    \label{tab:safetyrefs}
    \begin{tabular}{>{\raggedright}m{1.2cm} >{\centering\arraybackslash}m{1cm}| >{\centering\arraybackslash}m{0.5cm} >{\centering\arraybackslash}m{0.9cm} >{\centering\arraybackslash}m{0.5cm} >{\centering\arraybackslash}m{0.9cm}}
        \toprule
        \textbf{References} & \textbf{Fraction} & \multicolumn{2}{c}{\textbf{Temporal Safety}} & \multicolumn{2}{c}{\textbf{Spatial Safety}} \\
        \cmidrule(lr){3-4} \cmidrule(lr){5-6}
        & & \textbf{static} & \textbf{dynamic} & \textbf{static} & \textbf{dynamic} \\
        \midrule
        \textbf{Owning} & \textbf{\declownfrac\%} & \textbf{\safetyOwnTempstat}\% & \textbf{\safetyOwnTempdyn}\% & \textbf{\safetyOwnSpatstat}\% & \textbf{\safetyOwnSpatdyn}\% \\
        \quad - stack & \varsstackfrac\% & \safetyStackTempstat\% & \safetyStackTempdyn\% & \safetyStackSpatstat\% & \safetyStackSpatdyn\% \\
        \quad - heap & \varsheapfrac\% & 100.0\% & 0.0\% & 0.0\% & 100.0\% \\
        \quad - global & \varsglobalfrac\% & \safetyGlobalTempstat\% & \safetyGlobalTempdyn\% & \safetyGlobalSpatstat\% & \safetyGlobalSpatdyn\% \\
        \midrule
        \textbf{Borrowing} & \textbf{\declbrwfrac\%} & \textbf{\safetyRefTempstat}\% & \textbf{\safetyRefTempdyn}\% & \textbf{\safetyRefSpatstat}\% & \textbf{\safetyRefSpatdyn}\% \\
        \quad - mut. & \refbrwmutfrac\% & \safetyMutrefTempstat\% & \safetyMutrefTempdyn\% & \safetyMutrefSpatstat\% & \safetyMutrefSpatdyn\% \\
        \quad - immut. & \refbrwimmutfrac\% & \safetyImrefTempstat\% & \safetyImrefTempdyn\% & \safetyImrefSpatstat\% & \safetyImrefSpatdyn\% \\
        \midrule
        \midrule
        \textbf{Nullable} & \textbf{\optionallfrac}\% & \textbf{\safetyOptionTempstat}\% & \textbf{\safetyOptionTempdyn}\% & \textbf{\safetyOptionSpatstat}\% & \textbf{\safetyOptionSpatdyn}\% \\
        \midrule
        \textbf{DST} & \textbf{\dstallfrac}\% & \textbf{\safetyDstTempstat}\% & \textbf{\safetyDstTempdyn}\% & \textbf{0.0}\% & \textbf{100.0}\% \\
        \quad - string & \dststringfrac\% & \safetyStringTempstat\% & \safetyStringTempdyn\% & 0.0\% & 100.0\% \\
        \quad - buffer & \dstbufferfrac\% & \safetyBufferTempstat\% & \safetyBufferTempdyn\% & \safetyBufferSpatstat\% & \safetyBufferSpatdyn\% \\
        \quad - poly. & \dstpolyfrac\% & \safetyPolyTempstat\% & \safetyPolyTempdyn\% & \safetyPolySpatstat\% & \safetyPolySpatdyn\% \\
        \bottomrule
    \end{tabular}
\end{table}

\noindent\textbf{Breakdown of Different Types of References}.  Table \ref{tab:safetyrefs} summarizes the different types of smart references and data types that our users used, along with their frequency of usage. There are \todo{\declall} explicitly declared reference-like variables in the $31$ \todoadd{final} Rust translations in total. 
  Those variables can be classified into either owning references (\todo{\declownfrac}\%)\footnote{\todoaddcam{We consider smart pointers (e.g., \code{String} and \code{Vec}) as owning references.}} or borrowing references (\todo{\declbrwfrac}\%). 
    The owning references consist of references of stack data (\todo{\varsstackfrac}\%), heap data (\todo{\varsheapfrac}\%), and global data (\todo{\varsglobalfrac}\%). 
    Among borrowed references, \todo{\refbrwmutfrac}\% are mutable references, and \todo{\refbrwimmutfrac}\% are immutable references. 
\noindent\textbf{Spatial and Temporal Safety of References}.
All the references in safe Rust are strongly typed so that the Rust compiler can enforce certain safety invariants when using those types.
    For temporal safety guarantees, 
      most of the owning references (\todo{\safetyOwnTempstat}\%) and borrowing references (\todo{\safetyRefTempstat}\%) are compile-time checked, which are \emph{statically} proven to be free of temporal memory errors.
      This includes almost all of the stack and heap data references.
      The remaining are dynamic references, which involve either partial or full dynamic checks. 
      Typical types in the translations include \code{OnceCell<T>} (runtime check on the first write) and \code{Mutex<T>} (check on every code region of access).
      Most of the dynamic references are for global variables.
    For spatial safety,
      most references used (\safetyRefSpatdyn\%) may require runtime checks on access.
Two sub-categories of references, i.e., nullable references (\todo{\optionallfrac}\%) and references to dynamically-sized types (\todo{\dstallfrac}\%), affect abort handling and performance and are worth mentioning.

\noindent\textbf{Nullable References}.
Nullable references are typically represented by \code{Option\<T\>}, which piggybacks on static type safety to separate the case where a reference is \texttt{NULL} from when it is not.
It is often a zero-cost abstraction when \code{T} is a non-null reference, while \todoadd{preventing ungraceful aborts.}\tododel{ensuring that the Rust code will run {\em panic-free} (not raise runtime aborts).} Usage of nullable references forces the developer to specify how the code should handle null pointer deference, preventing the software from aborting ungracefully when memory safety is violated.
If the user does not want to specify how such exceptions should be handled, Rust gives a default way in which the compiler inserts null checks. They result in runtime panic on safety-violating inputs. It is explicitly reflected in the syntax (e.g., \code{unwrap(..)}). In the user translations, \todo{\optioncompfrac}\% of the accesses use such default null checks that may cause runtime panic on null pointer access, \todoadd{while the other $37\%$ use of \code{Option<T>} are \emph{panic-free} (not raising runtime aborts).} \tododel{Panic-free abstractions in Rust differ from many defenses which ungracefully abort without giving programmer control. Panic-free abstractions are used about $37\%$ of the time.}

\noindent\textbf{Dynamically Sized Types}.
Dynamically sized types (DSTs) are useful to support strings, buffers, as well as runtime polymorphism (e.g., \code{Box<dyn T>})~\cite{dst}. References of DSTs are typically ``fat'' pointers that store \tododel{more information than just a memory address.
The }additional information \todoadd{to }facilitates dynamic checks for spatial safety.
In our Rust translations, \todo{\dstallfrac}\% of the variables are DST references.\footnote{We count \code{String} and \code{Vec} as \todoaddcam{``references'' to} dynamically-sized types as well, even though those smart pointers themselves have fixed size.}

\begin{tcolorbox}[breakable,width=\linewidth,boxrule=1pt,top=0pt, bottom=0pt, left=1pt,right=1pt]
Temporal safety is achieved mostly statically (\todo{\safetyOwnTempstat}\%), whereas spatial safety is mostly through runtime checks.
\end{tcolorbox}

\noindent\textbf{Case Studies: Known Vulnerabilities Eliminated}. It is evident, even in our small-scale user study, that C to Rust translation directly addresses the root cause of memory safety vulnerabilities, namely insecure coding practices.
2 out of 8 of our benchmarks, \texttt{shoco} and \texttt{urlparser}, have $3$ known vulnerabilities in our chosen versions. The data compression library \texttt{shoco} has one spatial memory vulnerability (\href{https://nvd.nist.gov/vuln/detail/CVE-2017-11367}{CVE-2017-11367}) on the access of an array called \code{packs}. All users who translated this benchmark eliminated the spatial error and the out-of-bound is caught at runtime. Similarly, a heap-buffer overflow on a string buffer in \texttt{urlparser} is caught at runtime as well. The C \code{char*} pointers pointing to the string are lifted into 
\code{String}, \code{\&str}, or \code{\&String} with spatial safety guarantees. For temporal safety, there is a use-after-free (UAF) vulnerability~\footnote{The fix \href{https://github.com/jwerle/url.h/commit/752635e46be6b13ad045f7216a28417fdf533950\#diff-8eb10715daa9aa605b4e2c5e539f7b6641564a7fc2f2a329869ebc22187163baL101}{commit} changed the ownership from borrowing (line 100 of the left side) to owning (line 188-192 of the right side) in \code{url.c}.} in the \texttt{urlparser} C program when the input string to the parser does not live long enough before calling certain library APIs such as \code{url_data_inspect}. This bug is statically eliminated as the lifetime and borrowing rules in Rust forbid such code patterns. All participants created a copy of the borrowed input string that needed to live longer than the original, thereby eliminating the vulnerable pattern in the C code. More details on the case studies are in the Appendix~\ref{sec:apdxvuln}.

\begin{tcolorbox}[breakable,width=\linewidth,boxrule=1pt,top=0pt, bottom=0pt, left=1pt,right=1pt]
All the known memory safety vulnerabilities in C programs are eliminated in each one of the Rust translations.
\end{tcolorbox}

\subsection{Performance Comparison}
\label{sec:perf}

\begin{figure}[ht]
  \centering
  \begin{subfigure}[b]{0.49\linewidth}
    \includegraphics[width=\linewidth]{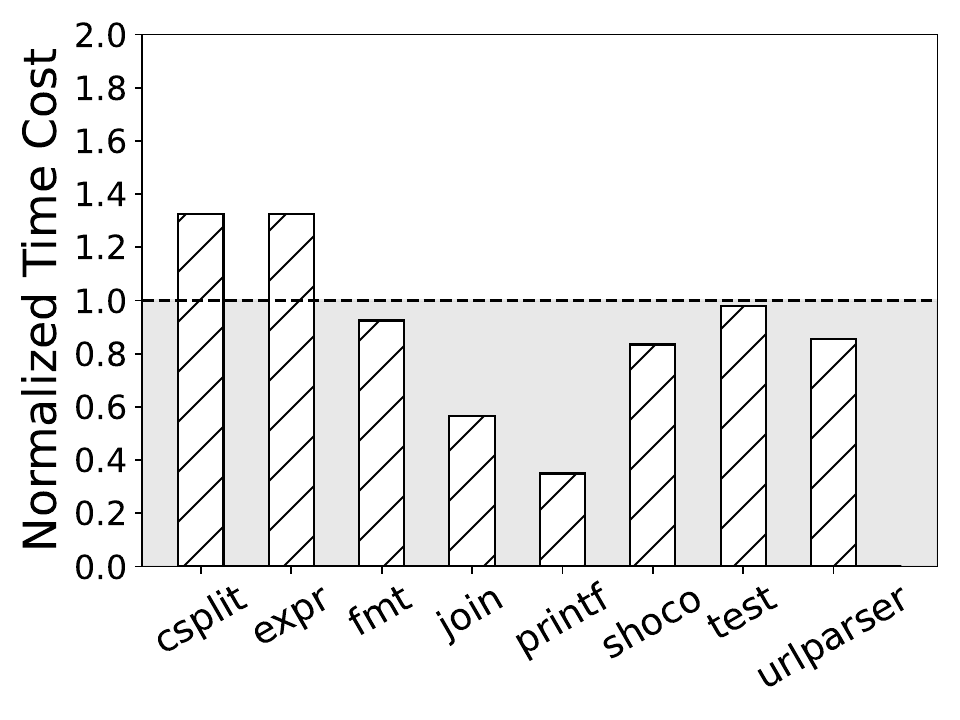}
    \caption{Compiled with \code{-O2}}
    \label{fig:perf_cmp2}
  \end{subfigure}
  \hfill
  \begin{subfigure}[b]{0.49\linewidth}
    \includegraphics[width=\linewidth]{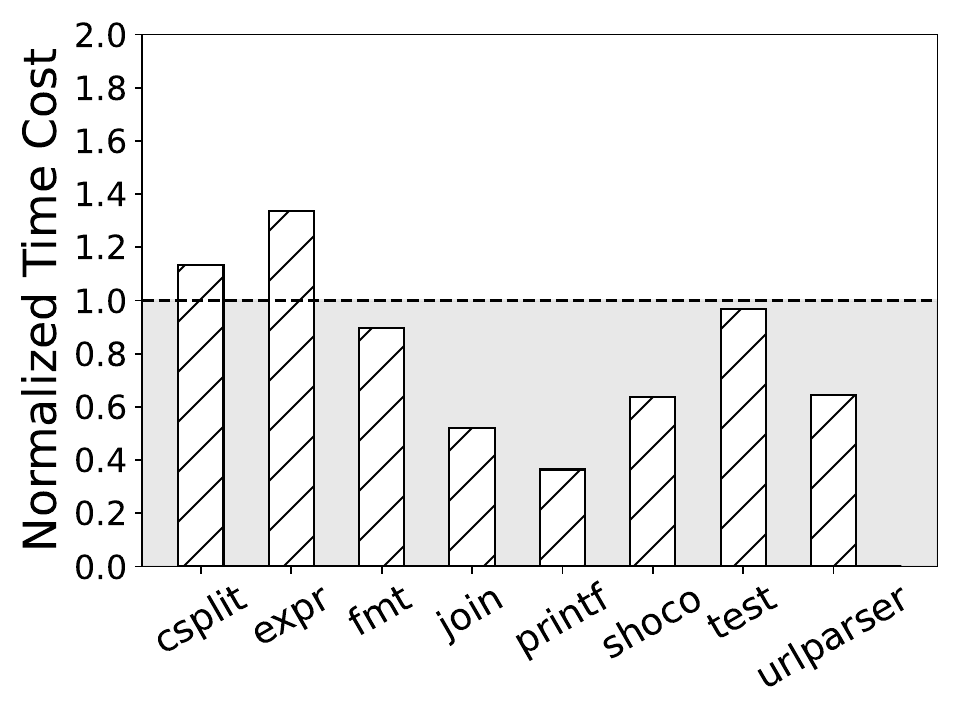}
    \caption{Compiled with \code{-O3}}
    \label{fig:perf_cmp3}
  \end{subfigure}
  \caption{Running time of the most similar Rust translation compared to the original C at different optimization levels. The time for the baseline C (dashed line) is normalized to 1.}
  \label{fig:perf_cmp_c_rs}
\end{figure}

We now turn to a comparison of performance overheads introduced when one migrates the same C code to Rust. An issue here is that our user-provided translations do not have the exact same behavior as the original C code on all inputs. Furthermore,
there are multiple candidate translations for each of $8$ C benchmarks to compare with, each with differences from the original C code. To deal with these, we choose the Rust program which is most similar to the corresponding C program on tests.
Specifically, we select the Rust translation with the highest test coverage as the main comparison baseline, tie-breaking in favor of the shorter translation. All the tests provided by all the participants for the C benchmarks are considered when computing the (line) coverage. This gives us $8$ Rust translations, one for each C benchmark. 

We report performance only on test cases for which the Rust code and corresponding C have the same output and both exit normally\footnote{We do not measure performance on unit tests for error handling.}, thereby eliminating unfair comparisons.\tododel{ of dissimilar behavior.}

The running time is calculated between the entry and exit of the \code{main} function of the programs to exclude differences in the load and initialization time\footnote{We also measure end-to-end performance. Details are in the Appendix~\ref{sec:e2eperf}.}. We report the average over \todo{70} repeated trials, removing the first 10 rounds, which warm up hardware and OS caches, as well as the fastest and slowest 10 rounds of the remaining. The ratio of running time Rust compared to the corresponding C is given in Fig. \ref{fig:perf_cmp_c_rs}. 
We have that for \todoadd{$5$} out of the $8$ C benchmarks, their most similar Rust translations run faster. $6$
out of $8$ have corresponding Rust translations well below $10\%$ overheads, while the remaining $2$ have about $10\%$
and $40\%$ overhead at \code{-O3} optimization level.

It is worth noting that our performance analysis is post-hoc. We did not ask users to measure or optimize for performance. We only specified correctness and safety as the objectives. 

\begin{tcolorbox}[breakable,width=\linewidth,boxrule=1pt,top=0pt, bottom=0pt, left=1pt,right=1pt]
For Rust translations most similar to the original C code, the overhead is mostly within $20\%$ and Rust is often faster.
\end{tcolorbox}

\subsection{\todoadd{Examples of Security-Enhancing Code Patterns}}
\label{sec:safety2}

\todoadd{Certain code patterns in C are considered unsafe and are forbidden in safe Rust. However, there are often safe abstractions in Rust designed to achieve similar goals at low cost.
We focus on two kinds of patterns present in our C programs, including mutable globals and unions, 
to analyze how users translate such code patterns into safe Rust.}

\noindent\todoadd{\textbf{Mutable Globals.} Mutable global variables in C are prone to be corrupted and exploited in real-world attacks due to their extensive lifetime across many functions and relatively predictable address~\cite{li2023hybrid,practicaldoa}.
Besides being a convenient target for attackers, mutable globals are also sources of various hard-to-find temporal memory errors, and are thus discouraged by various coding guidelines~\cite{globalgoogle,globallinux,globalcpp}. 
Safe Rust forbids plain mutable globals because they violate the AXM principle even in single-threaded programs. Otherwise, different call frames can obtain mutable references to globals with overlapping lifetimes, leading to memory errors and incorrect optimizations in Rust.
To translate C code using globals into safe Rust, we see that users devised several strategies using safe Rust abstractions, which are listed below:
\begin{itemize}
    \item[A)] \emph{From global to locals.} A performant strategy is to identify where the global is used first and last in the program. Then, one can replace the original global variable with a local (stack or heap) object spanning the lifetime of actual use. A reference can be passed into functions that need them. When multiple mutable globals are transformed this way, they can be grouped into a single \code{struct} object for better performance and to reduce code bloat. 
    One reference to the grouped \code{struct} object with the original globals as fields at different offsets is sufficient. This optimizes for performance as it lowers the cost of parameter passing by reference for many moved globals with similar lifetimes of use in the original C code. 
    \item[B)] \emph{Dynamic references.} The fallback strategy is to create dynamic references, declared either as thread locals or true globals in Rust. For thread-local variables, single-thread synchronization (e.g., \code{RefCell}) is still needed~\cite{rusttls}, but multi-thread synchronization (e.g., \code{Mutex}) is not necessary. Such references incur some runtime costs. 
    \item[C)] \emph{Atomic integer types.}
    For mutable global variables that are integers, dynamic references are not needed if declaring them as \code{Atomic}. For \code{Atomic} types, special APIs can mutate their values without mutable references to them.
\end{itemize}
}

$19$ participants chose to move the globals to locally referenced objects. $11$ of them also grouped multiple globals into an aggregate \code{struct} object. $3$ users kept mutable globals as globals or thread-local variables through dynamic references. $4$ used atomics for mutable globals. 
\ifdefined\confver
\todoaddcam{\extendappendix}
\fi

\noindent\todoadd{\textbf{C Unions.} Unions in C allow overlapping objects of incompatible types.
However, programmers are responsible for ensuring that operations on unions are type safe. 
Otherwise, type safety violations can lead to both spatial and temporal memory errors~\cite{effectivesan}.
In safe Rust, unions are not allowed, but various abstractions can be used to achieve similar functionality.
$2$ of the 8 C programs in our benchmarks use unions. 
Users translated C unions depending on different use cases:
\begin{itemize}
    \item[A)] \emph{Zero-cost type casting (type punning~\cite{type-punning}).} The string compression library \texttt{shoco} uses union to access integers as raw bytes. Its behavior depends on the endianness of the target architecture. Some users translate such code into safe Rust API calls (\code{to(from)_le(be)_bytes}) that convert between primitives and constant-size byte arrays.
    Note that the conversion is zero-cost since those APIs will be optimized away during compilation. They can also be used together with conditional compilation supported in safe Rust to match the endianness of the architecture. 
    \ifdefined\confver
    \todoaddcam{Examples are provided in Appendix D in the extended version~\cite{userstudy} of this paper.}
    \else
    Please see Fig.~\ref{fig:unionpun} in the Appendix~\ref{sec:apdxpattern} for examples.
    \fi
    \item[B)] \emph{Sum type (variant record).} Another C program \texttt{expr} uses unions to implement variant records (sum types). For such use cases, some users translate them into \code{enum} and access them using statically-checked \code{match} statements.\tododel{in safe Rust, which provides an identical abstraction.} Rust \code{enum} is memory efficient, and the memory occupied by a Rust \code{enum} depends on its largest discriminant, similar to unions in C.
    \ifdefined\confver
    \todoaddcam{Please see Appendix D in the extended version~\cite{userstudy} for examples.}
    \else
    An example (Fig. \ref{fig:unionenum}) is in the Appendix~\ref{sec:apdxpattern}.
    \fi
\end{itemize}
For relevant use cases of unions, $3$ users used zero-cost type casting APIs in safe Rust, and $2$ users used Rust enums. 
Another $3$ users did not use those abstractions but emulated the C unions using structs and methods with higher overhead.}

\begin{tcolorbox}[breakable,width=\linewidth,boxrule=1pt,top=0pt, bottom=0pt, left=1pt,right=1pt]
\todoadd{
Users often translate known unsafe C code patterns to equivalent statically-checked low-cost safe Rust code.}
\end{tcolorbox}

\section{The Gap In Functional Correctness}
\label{sec:testdiff}

As mentioned in Section \ref{sec:userstudy}, each participant submitted their final translation and tested it with test cases they created.
Most participants created tests that covered more than 85\% of the original C program and reported that their final Rust translations passed their tests.
We analyze how many behavioral differences are missed by tests self-created by participants. 
We employ automated fuzz-testing to check for behavioral differences between Rust and the corresponding C source. We use \texttt{AFL++}~\cite{aflplusplus} to generate tests for each of the $8$ C programs for \todoadd{$1$ hour} per program. We then sample $300$ distinct tests per program, except for programs with less than $300$ tests available, in which case we took all tests\footnote{$3$ programs had fewer than $300$ tests: \texttt{shoco} ($200$ tests), \texttt{csplit} ($211$ tests), and \texttt{urlparser} ($22$ tests)}.
On these tests, we compare each of the $31$ Rust programs to its corresponding C program. We omit minor \todoadd{non-semantic} discrepancies 
\todoadd{such as the error message format }when computing the difference.

We find that {\em none of the 31 translations is fully equivalent} to the original C code. 
Fig.~\ref{fig:test_count} shows that \todo{\cmpFuzzMinAllfail}\%-\todo{\cmpFuzzMaxAllfail}\% (\todo{\cmpFuzzAvgAllfail}\% on average) of the fuzz tests exhibit a discrepancy across different Rust translations of the same C program.

\begin{figure}[t]
  \centering
    \includegraphics[width=0.8\linewidth]{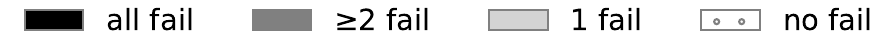}
    \begin{subfigure}[b]{0.49\linewidth}
    \includegraphics[width=\linewidth]{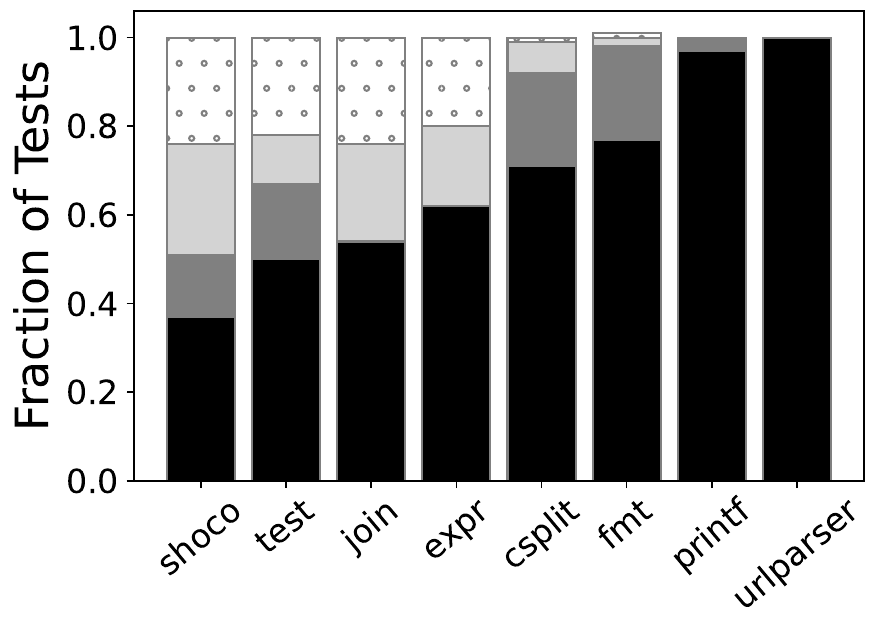}
    \caption{Greybox fuzzing tests}
    \label{fig:common_fail_fuzz}
  \end{subfigure}
  \hfill
  \begin{subfigure}[b]{0.49\linewidth}
    \includegraphics[width=\linewidth]{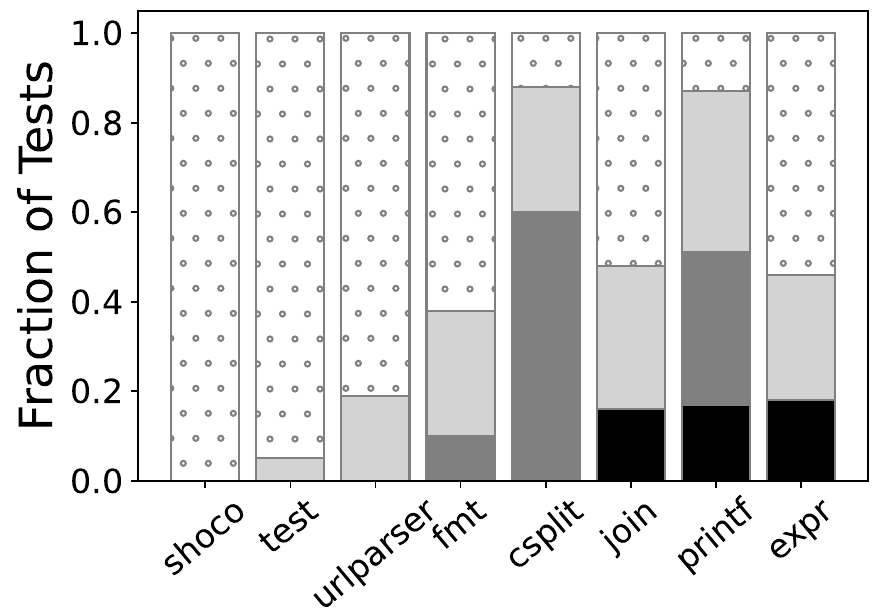}
    \caption{User-provided tests}
    \label{fig:common_fail_union}
  \end{subfigure}
  \caption{Rust translations of the same C program all fail on a large fraction of fuzz tests (Subfig (a)) and all pass on a large fraction of user-provided tests (Subfig(b)).} 
  \label{fig:test_count}
\end{figure}

\noindent\textbf{Many translations fail the same way}. 
Given that all the translations have behavioral differences from the corresponding C code, we examine whether their failures are correlated. We count how many translations for the same C program fail on the {\em same} test case. 
Fig. \ref{fig:test_count} shows that nearly all Rust translations  fail on about the same $38\%$ tests on \texttt{shoco},
the same $50\%$ on \texttt{test}, and so on.
We investigated some of the tests that all translations failed, finding that they are often corner cases involving C library calls that are tedious to emulate in safe Rust, and hence missed by all translations. An illustrative example is provided in the Appendix~\ref{sec:applogic} for interested readers.

\begin{figure}[t]
  \centering\footnotesize

  \definecolor{lightgreen}{RGB}{73,200,73}
\definecolor{lightgray}{RGB}{221,221,221}
\definecolor{lightbrown}{RGB}{231,181,59}
\definecolor{lightred}{RGB}{255,122,133}

\begin{tikzpicture}
    \begin{axis}[
        xbar stacked,   
        xmin=0, xmax=100, 
        area style,
        ytick=data,    
        yticklabels={test-4,test-3,test-2,test-1,,expr-2,expr-1,,printf-3,printf-2,printf-1,,join-2,join-1,,csplit-5,csplit-4,csplit-3,csplit-2,csplit-1,,fmt-4,fmt-3,fmt-2,fmt-1,,shoco-6,shoco-5,shoco-4,shoco-3,shoco-2,shoco-1,,urlparser-5,urlparser-4,urlparser-3,urlparser-2,urlparser-1},
        tick align=outside,
        axis y line*=none, 
        axis x line*=bottom,
        xlabel={(percentage of fuzzer-generated tests)}, 
        xmajorgrids = true,
        bar width=0.7mm,
        y=1.3mm,
        enlarge y limits={abs=0.625},
        tick label style={font=\tiny}
    ]

\addplot [fill=lightgreen,draw=lightgreen] coordinates {(47.33, 0) (44.33, 1) (31.0, 2) (25.67, 3) (0, 4) (37.67, 5) (21.0, 6) (0, 7) (2.33, 8) (1.67, 9) (0.33, 10) (0, 11) (35.67, 12) (34.33, 13) (0, 14) (27.96, 15) (18.96, 16) (11.37, 17) (9.95, 18) (6.64, 19) (0, 20) (24.33, 21) (14.33, 22) (8.0, 23) (3.67, 24) (0, 25) (61.67, 26) (61.67, 27) (57.33, 28) (57.33, 29) (45.0, 30) (28.33, 31) (0, 32) (0.0, 33) (0.0, 34) (0.0, 35) (0.0, 36) (0.0, 37) };
\addplot [fill=lightgray,draw=lightgray] coordinates {(50.0, 0) (50.0, 1) (50.0, 2) (50.0, 3) (0, 4) (50.67, 5) (50.67, 6) (0, 7) (82.33, 8) (82.33, 9) (82.33, 10) (0, 11) (31.67, 12) (31.67, 13) (0, 14) (45.5, 15) (45.5, 16) (45.5, 17) (45.5, 18) (45.5, 19) (0, 20) (11.0, 21) (66.0, 22) (0.0, 23) (54.67, 24) (0, 25) (0, 26) (0, 27) (0, 28) (0, 29) (0, 30) (0, 31) (0, 32) (50.0, 33) (50.0, 34) (50.0, 35) (50.0, 36) (50.0, 37) };
\addplot [fill=lightbrown,draw=lightbrown] coordinates {(2.0, 0) (0.0, 1) (0.0, 2) (0.0, 3) (0, 4) (0, 5) (0, 6) (0, 7) (3.0, 8) (0.0, 9) (0.0, 10) (0, 11) (0.0, 12) (0.0, 13) (0, 14) (10.43, 15) (0.0, 16) (0.95, 17) (2.37, 18) (0.0, 19) (0, 20) (0.0, 21) (3.0, 22) (0.33, 23) (3.33, 24) (0, 25) (30.67, 26) (30.67, 27) (30.67, 28) (30.67, 29) (31.33, 30) (28.33, 31) (0, 32) (0.0, 33) (0.0, 34) (4.55, 35) (36.36, 36) (40.91, 37) };
\addplot [fill=lightred,draw=lightred] coordinates {(0.67, 0) (5.67, 1) (19.0, 2) (24.33, 3) (0, 4) (11.67, 5) (28.33, 6) (0, 7) (12.33, 8) (16.0, 9) (17.33, 10) (0, 11) (32.67, 12) (34.0, 13) (0, 14) (16.11, 15) (35.55, 16) (42.18, 17) (42.18, 18) (47.87, 19) (0, 20) (64.67, 21) (16.67, 22) (91.67, 23) (38.33, 24) (0, 25) (7.67, 26) (7.67, 27) (12.0, 28) (12.0, 29) (23.67, 30) (43.33, 31) (0, 32) (50.0, 33) (50.0, 34) (45.45, 35) (13.64, 36) (9.09, 37) };

    \end{axis}
\end{tikzpicture}
  \caption{Behavioral differences across $31$ translations with tests from greybox fuzzing. Legends from left to right mean ``\textit{equivalent behaviors}'' (\textcolor{lightgreen}{green}), ``\textit{I/O enc. errors}'' (\textcolor{mydarkgray}{gray}), ``\textit{Runtime safety aborts}'' (\textcolor{mybrown}{brown}), and ``\textit{Logical differences}'' (\textcolor{lightred}{red}).}
  \label{fig:diffbreakdown}
\end{figure}

\noindent\textbf{Breakdown of Behavioral Differences}. The behavioral differences can be classified into three categories, including:
\begin{itemize}
    \item[A)] \emph{I/O encoding/decoding errors.} Standard Rust APIs have different I/O formatting and encoding behavior from similar C APIs. For example, the input arguments and string APIs often assume UTF-8 encoding, which might abort the program if the string buffer is not valid UTF-8.
    \item[B)] \emph{Runtime safety aborts.} The Rust has runtime checks about spatial errors and null pointer dereferences that safely abort, while C may continue arbitrarily and crash.
    \item[C)] \emph{Logical differences.} Both C and Rust can finish the execution and normally exit, but the output is different.
\end{itemize}
Fig. \ref{fig:diffbreakdown} shows the distribution of category of differences for each of the $31$ translations. 
Nearly all translations (\todo{27}/\todo{31}) behave differently on more than half of the newly created tests from fuzzing. 
Logical differences (category C) exist across all translations, highlighting the gap in functional equivalence.
I/O encoding/decoding errors (category A) are frequent in general, while safety-violating inputs (category B) are more frequent in translations of \texttt{shoco} and \texttt{urlparser}. Note that Category B cases do not lead to undefined behavior in Rust due to
runtime safety checks, but exhibit behavioral differences to C. These differences are to be expected---Rust is eliminating potentially unsafe (undefined) behavior from untamed C code.

\emph{Differences in I/O.}
Although the gap between the translations and full functional equivalence is non-negligible and common, we find that some of the differences may not matter for many intended usage scenarios.
One example is that the differences in I/O encoding/decoding behaviors are a major source of behavioral differences, as shown in the gray bars in Fig. \ref{fig:diffbreakdown}. 
Such differences may not matter if the intended inputs to the programs are valid UTF-8 strings.
The UTF-8 encoding is widely assumed in the Rust standard library on string-related APIs.
Those APIs provide convenient ways to work with UTF-8 encoded strings compared with other encodings.
If the intended use case requires processing other encodings, the same data types and APIs may no longer be useful. Rewriting of related API functions or external emulation would be necessary, leading to a more verbose translation.

\emph{Panic-free Translations.} $7$ out of $8$ C benchmarks have at least one user-provided Rust translation that is panic-free\footnote{Excluding the abort due to I/O encoding/decoding errors.}. The \texttt{urlparser} benchmark is an example where we have $2$ translations that do not have any runtime safety aborts (brown bar in Fig. \ref{fig:diffbreakdown}), while the remaining $3$ translations have panics.
On the contrary, the $2$ panic-free translations avoid all use of \code{Option<T>::unwrap()}. 
These differences across Rust translations are only artifacts of how different users handle safety aborts, arising out of undefined behavior in C.

\begin{tcolorbox}[breakable,width=\linewidth,boxrule=1pt,top=0pt, bottom=0pt, left=1pt,right=1pt]
None of the Rust translations is fully equivalent to the corresponding C, but some differences are due to Rust programmers handling safety aborts differently. Several logical differences and incompatibilities in the external I/O remain.
\end{tcolorbox}

\section{Efficacy of State-of-the-art Automatic Tools}
\label{sec:existingcompare}
\newcolumntype{N}{>{\centering\arraybackslash}m{0.7cm}}
\newcolumntype{C}{>{\centering\arraybackslash}m{0.58cm}}
\newcolumntype{S}{>{\centering\arraybackslash}m{0.5cm}}
\begin{table*}[ht]
    \centering
    \caption{
      Comparison of Existing Tools on Our Benchmarks
    }
    \label{tab:existing}

\begin{minipage}{\textwidth}
\begin{tabular}{|l||N|CC|C||N|SS|C||N|SS|C||N|SS|C|}
    \hline
     \backslashbox[16mm]{Prog.}{Tools} 
    & \multicolumn{4}{c||}{\laertes} & \multicolumn{4}{c||}{\crown} & \multicolumn{4}{c||}{\florine} & \multicolumn{4}{c|}{\toolvert} \\ \cline{2-17} 
    & \multirow{2}{*}{Compile} & \multicolumn{2}{c|}{Safe Code (\%)} & \multirow{2}{*}{\parbox{0.6cm}{Passed\newline Tests}} & \multirow{2}{*}{Compile} & \multicolumn{2}{c|}{Safe Code (\%)} & \multirow{2}{*}{\parbox{0.6cm}{Passed\newline Tests}} & \multirow{2}{*}{Compile} & \multicolumn{2}{c|}{Safe Code (\%)} & \multirow{2}{*}{\parbox{0.6cm}{Passed\newline Tests}} & \multirow{2}{*}{Compile} & \multicolumn{2}{c|}{Safe Code (\%)} & \multirow{2}{*}{\parbox{0.6cm}{Passed\newline Tests}} \\ \cline{3-4}\cline{7-8}\cline{11-12}\cline{15-16} 
    &  & \#Line & \#Ref. &  &  & \#Line & \#Ref. &  &  & \#Line & \#Ref. &  &  & \#Line & \#Ref. &  \\ \hline
    \texttt{csplit} & \ding{51} & \safetyLaertesCsplitLine\% & \safetyLaertesCsplitPtrDecl\% & \correctLaertesCsplit\% & \ding{51} & \safetyCrownCsplitLine\% & \safetyCrownCsplitPtrDecl\% & \correctCrownCsplit\% & \ding{55} & \safetyFlourineCsplitLine\% & \safetyFlourineCsplitPtrDecl\% & - & \ding{55} & \safetyVertCsplitLine\% & \safetyVertCsplitPtrDecl\% & - \\ \hline
    \texttt{expr} & \ding{51} & \safetyLaertesExprLine\% & \safetyLaertesExprPtrDecl\% & \correctLaertesExpr\% & \ding{55} & \safetyCrownExprLine\% & \safetyCrownExprPtrDecl\% &- & \ding{55} & \safetyFlourineExprLine\% & \safetyFlourineExprPtrDecl\% & - & \ding{55} & \safetyVertExprLine\% & \safetyVertExprPtrDecl\% & - \\ \hline
    \texttt{fmt} & \ding{55} & \safetyLaertesFmtLine\% & \safetyLaertesFmtPtrDecl\% & - & \ding{51} & \safetyCrownFmtLine\% & \safetyCrownFmtPtrDecl\% & \correctCrownFmt\% & \ding{55} & \safetyFlourineFmtLine\% & \safetyFlourineFmtPtrDecl\% & - & \ding{55} & \safetyVertFmtLine\% & \safetyVertFmtPtrDecl\% & - \\ \hline
    \texttt{join} & \ding{51} & \safetyLaertesJoinLine\% & \safetyLaertesJoinPtrDecl\% & \correctLaertesJoin\% & \ding{55} & \safetyCrownJoinLine\% & \safetyCrownJoinPtrDecl\% & - & \ding{55} & \safetyFlourineJoinLine\% & \safetyFlourineJoinPtrDecl\% & - & \ding{55} & \safetyVertJoinLine\% & \safetyVertJoinPtrDecl\% & - \\ \hline
    \texttt{printf} & \ding{51} & \safetyLaertesPrintfLine\% & \safetyLaertesPrintfPtrDecl\% & \correctLaertesPrintf\% & \ding{51} & \safetyCrownPrintfLine\% & \safetyCrownPrintfPtrDecl\% & \correctCrownPrintf\% & \ding{55} & \safetyFlourinePrintfLine\% & \safetyFlourinePrintfPtrDecl\% & - & \ding{55} & \safetyVertPrintfLine\% & \safetyVertPrintfPtrDecl\% & - \\ \hline
    \texttt{test} & \ding{51} & \safetyLaertesTestLine\% & \safetyLaertesTestPtrDecl\% & \correctLaertesTest\% & \ding{51} & \safetyCrownTestLine\% & \safetyCrownTestPtrDecl\% & \correctCrownTest\% & \ding{55} & \safetyFlourineTestLine\% & \safetyFlourineTestPtrDecl\% & - & \ding{55} & \safetyVertTestLine\% & \safetyVertTestPtrDecl\% & - \\ \hline
    \texttt{shoco} & \ding{51} & \safetyLaertesShocoLine\% & \safetyLaertesShocoPtrDecl\% & \correctLaertesShoco\% & \ding{51} & \safetyCrownShocoLine\% & \safetyCrownShocoPtrDecl\% & \correctCrownShoco\% & \ding{55} & \safetyFlourineShocoLine\% & \safetyFlourineShocoPtrDecl\% & - & \ding{55} & \safetyVertShocoLine\% & \safetyVertShocoPtrDecl\% & - \\ \hline
    \texttt{urlparser} & \ding{51} & \safetyLaertesUrlparserLine\% & \safetyLaertesUrlparserPtrDecl\% & \correctLaertesUrlparser\% & \ding{51} & \safetyCrownUrlparserLine\% & \safetyCrownUrlparserPtrDecl\% & \correctCrownUrlparser\% & \ding{55} & \safetyFlourineUrlparserLine\% & \safetyFlourineUrlparserPtrDecl\% & - & \ding{55} & \safetyVertUrlparserLine\% & \safetyVertUrlparserPtrDecl\% & - \\ \hline
\end{tabular}
\end{minipage}
\end{table*}

We evaluate $4$ state-of-the-art C to Rust translation tools on our C benchmarks to analyze whether their approach is effective and, when not, where the immediate gaps are. 

\subsection{Compiler-based Tools}
Both \laertes~and \crown~are compiler-based systems that post-process the unsafe Rust code produced by the \texttt{c2rust} compiler~\cite{c2rust}. 
Our results, shown in Table \ref{tab:existing}, confirm that while both tools 
\todoaddcam{can often produce Rust programs that compile,}
\todoadd{the generated code is still largely unsafe and far from safe Rust.}
We summarize their missing features next.

\todoadd{
\emph{Fraction of Raw Pointers.} 
The fraction of raw pointers that can be lifted by \laertes~and \crown~ is limited. For \laertes, Most of the data references (\todo{\unsafetyLaertesPtrDeclAvg}\% on average) are still unsafe raw pointers\footnote{\todoaddcam{Excluding data references in helper functions injected by \laertes.}}. For \crown, the majority of the references are also raw pointers, and safe Rust references account for less than \todo{\safetyCrownPtrDeclAvg}\% averaged over all the benchmarks. 
}

\emph{Lifted Safe Rust Data Types.} Both of the translation tools have limited support for higher-level data types. 
Both \laertes and \crown are limited to three types of Rust references that are closely related to C pointers. The three types include \code{Option<\&T>} (similar to \code{const T*} in C), \code{Option<\&mut T>} (similar to \code{T*}), and \code{Option<Box<T>>} (owning \code{T*}) where \code{T} is a C-compatible type.
Most of the higher-level smart references that participants frequently use are out of the scopes of those tools, including \code{String}, \code{\&str}, \code{Vec<T>}, \code{BufRead}, and so on, which are shown in Fig. \ref{fig:typemap}. 

\emph{Handling of C Library Calls.}
C library calls turn into FFI calls from Rust to C in the translation of both \laertes~and \crown. 
C library functions generally do not have direct safe Rust mappings, and the translation is often not straightforward. 
Calling C functions through FFI is easy, but such handling of library calls permits unsafety.
Such operations are only allowed in unsafe Rust. At the same time, C APIs often require raw pointers as parameters, which may also constrain related Rust code to operate on raw pointers.
In contrast, users in our study addressed the API translation problem by utilizing safe Rust code in creative ways, as shown previously in Fig. \ref{fig:fig_api_mapping}. 

\emph{Dealing with Aliasing and Lifetimes.} 
Aliasing references in Rust can frequently violate borrowing rules and thus can be challenging to translate.
\todoadd{
\laertes~and \crown~ aim to lift pointers to references while preserving the low-level control flow and data flow of the C program. The strategies highlighted in our work, for instance in Section~\ref{sec:aliasing}, are not used by these tools.
These prior tools do not aim to eliminate unsafe C code patterns such as mutable globals or unions that may be present in the \texttt{c2rust} compiler output used as their first stage.}

\emph{Memory Safety.} The end result of the code produced from prior compiler-based tools \todoadd{might be safer than C, but is not guaranteed to have full memory safety.}
Recall that there are $3$ known vulnerabilities in our C benchmarks. 
\todoadd{
The translation resulting from these tools eliminates $1$ of these vulnerabilities, a read overflow of a global array, as shown in Fig. \ref{fig:case_study} in the Appendix~\ref{sec:apdxvuln}. 
The original global array declaration in C is syntactically translated into an equivalent global array declaration in Rust. Access to such an array (rather than via a raw pointer) is bounds-checked automatically in Rust.
The remaining $2$ vulnerabilities persist in the Rust translation produced by both these tools, one of which is a spatial violation and the other temporal. Raw C pointers (not integers) are involved in the unsafe code in the $2$ cases, and \laertes~and \crown~ are not yet able to lift them to safe Rust references.
}

\begin{tcolorbox}[breakable,width=\linewidth,boxrule=1pt,top=0pt, bottom=0pt, left=1pt,right=1pt]
State-of-the-art compiler-based C to Rust tools output translations that extensively use unsafe Rust while rigidly retaining many structural similarities to the original C code.
\end{tcolorbox}

\subsection{LLM-based Tools}
There are $2$ recent tools, \florine~and \toolvert, that are state-of-the-art for C to Rust translation utilizing large language models (LLMs). We report on our experience in running them on our $8$ benchmark programs.
\florine~explores $4$ test-driven repair strategies in concert with LLMs. 
\toolvert~uses a different approach. It creates two translations of the given C program, one is unsafe Rust decompiled from WASM, and the other is an LLM-\todo{generated} safe Rust code. Then, it uses fuzzing and, if needed, model-checking techniques to find tests that exhibit differences and subsequently run an automatic repair. The work on \toolvert claims that if their tool terminates normally, the resulting translation is expected to be functionally equivalent to the C code. 
\todoaddcam{For evaluating \florine, we choose their most stable configuration with GPT-4\footnote{\todoaddcam{We use the model version gpt-4-0125.}} as the LLM backend. For evaluating \toolvert, we use Claude\footnote{\todoaddcam{We use Claude 3.5 (claude-3-5-sonnet-20240620), which is the best available version of Claude at the time of our evaluation. 
}
}, which is reported to be the best-performing LLM backend for their artifact.}
\todoadd{We observe that code generated from these LLM-based tools is more idiomatic and uses safe Rust abstractions more often than compiler-based tools, and thus can serve the goal of useful translation aids for human developers on code snippets.} However, neither tool generates a runnable translation for any of the C programs we consider, as shown in Table \ref{tab:existing}.
This is because both tools have certain structural assumptions on the C code given to them as input.
Specifically, both tools expect that the C program can be decomposed into components such that each can be \emph{independently} translated and tested, since present LLMs work reasonably well on small code fragments.

The decomposition of C programs into such components and merging of their Rust translations are not automated by the tools. We encountered many difficulties when trying to emulate the decomposition strategies described in their respective works. These difficulties point to a broader technical challenge that may be of independent interest for further research.

\emph{Decomposition Failures.}
$6$ out of $8$ of our C benchmarks are real-world standalone programs, and the remaining $2$ are libraries. 
Functions in standalone programs are connected by a call graph\tododel{that captures their invocation order}. We find it difficult to decompose such functions into independently testable modules using the strategies described in the works of \florine or \toolvert. 
For example, if a function $f$ calls $g$, then the component containing $f$ is required to contain $g$ since it is a dependency. 
\tododel{The function $g$ can be in a component by itself.}This implies that on our standalone C program benchmarks, the function \code{main} depends on all other functions and the component containing it \tododel{must contain all other functions. This }includes nearly the whole program, making it too large to obtain repairable translations from LLMs. The $2$ library benchmarks are marginally better since there is no \code{main} function, but the challenge persists in part here as well. 

We note that the decomposition strategy proposed in \florine or \toolvert can produce multiple translations of the same function
that are inconsistent and cannot be merged into one. Say we have two components, one containing function $\{f,g\}$ and another containing function $\{h,g\}$. The two components are translated independently and thus multiple Rust translations of the function $g$ are obtained. We often find that these translations have conflicting type definitions and incompatible types that are difficult to merge into a single translation\tododel{ for $g$}.

We are not aware of better ways to decompose our C programs in components small enough to feed to \florine or \toolvert. We fed the whole program to these tools and the translated Rust code does not compile. We then experimented with several versions manually to best split each function in a separate component, while including only a minimal number of dependencies, such that the size of component is small enough to work with LLMs. We are able to produce Rust translations reported in Table~\ref{tab:existing} with some compilable components, \todoadd{but unable to merge them back into a single compilable program}.

\begin{tcolorbox}[breakable,width=\linewidth,boxrule=1pt,top=0pt, bottom=0pt, left=1pt,right=1pt]
State-of-the-art LLM-based C to Rust translators produce idiomatic safe Rust snippets, but not safe Rust programs that compile. Existing tools share a common challenge in devising workable decomposition strategies for long C code. 
\end{tcolorbox}

\subsection{Do existing LLMs help in user-provided translation?}
We revisit whether LLMs, taken standalone, are helpful to users who followed their own translation strategies. Recall that we placed no restrictions on the participants\tododel{ of the user study} to employ external tools. We asked our participants to specify which \tododel{external }tools they used and provide qualitative feedback on their experience.
In their feedback, $31$ of the $33$ participants reported that they tried to use LLMs for assistance\tododel{to assist with code translation}.
$14$ users mentioned that LLMs are helpful indirectly in the translation process, including tasks such as explaining the C code and suggesting Rust data types and APIs. However, most of the participants ($20$/$31$) reported that the code generated by LLMs is error-prone and hard to debug.
2 users reported that they abandoned LLMs for direct code translation and translated manually from scratch.

\section{\todoadd{Extensibility of Findings to Real-world Code}}
\label{sec:experts}

\tododelsec{Readers might wonder whether our findings are specific to our chosen programs or our user group. We check if these findings apply to real-world Rust programs by conducting a post-hoc analysis of a mature open-source Rust project \href{https://github.com/uutils/coreutils}{\texttt{uutils/coreutils}}.}
\todoadd{To test whether our findings generalize beyond our chosen programs or our user group, we conduct a post-hoc analysis of a mature open-source Rust project \href{https://github.com/uutils/coreutils}{\texttt{uutils/coreutils}}
\footnote{\todoadd{We download a version on Sept. 26th, 2024 with commit hash \href{https://github.com/uutils/coreutils/tree/a0d258d3f29cbe6b6714b4758554dba0e84264c8}{a0d258d}.}} (\texttt{uutils}),} 
\todoadd{which mirrors the \href{https://www.gnu.org/software/coreutils/}{GNU coreutils} written in C. 
The \texttt{uutils} project aims for compatibility with GNU coreutils and passing the same GNU test suite.
The \texttt{uutils} Rust repository has $17.6k$ Github stars and $5$ of its most active contributors have Rust experience of more than $2$ years at the time of this writing.
We examine $6$ \texttt{uutils} Rust programs that share names and functionality with the $6$ \texttt{BSDCoreUtils} programs in our benchmarks.
Similar to our user study, we compare those Rust programs to their C counterparts in GNU coreutils.}

\noindent\todoadd{\textbf{Program Dissimilarity.} Recall that the Rust translations in our user study are dissimilar from the C source (Sec.~\ref{sec:lifting}). We find that a larger dissimilarity between the structure of \texttt{uutils} Rust programs and their corresponding GNU C programs compared with our user study is observed.
We manually checked $142$ and $165$ C and Rust functions, respectively, across the $6$ programs. We were only able to find $9$ pairs of functions semantically similar, and they were all relatively small.\footnote{\todoadd{Such large dissimilarity may be due to the fact that \texttt{uutils/coreutils} is re-implemented in Rust from scratch to avoid license issues.}}}

\noindent\todoadd{\textbf{Security-Enhancing Code Patterns.}
The code patterns involving mutable globals and unions in C are changed in Rust programs in our user study (Sec. \ref{sec:safety2}). It is the same for the \texttt{uutils} project.
The original GNU C programs also have many global variables (on average $13.2$ globals per program), and
the corresponding Rust programs in \texttt{uutils} avoid global variables almost completely. These Rust programs group variables into \code{struct}-typed objects (strategy A for globals in Sec. \ref{sec:safety2}), more often than in our study.
One program \texttt{expr}, which involves unions in C, is implemented with \code{Enum}s in safe Rust.}

\noindent\todoadd{\textbf{Semantic Data Type Lifting.} In our study, users lift raw pointers into various safe Rust data types, as shown previously in Fig.~\ref{fig:typemap}. We find that Rust data types used in \texttt{uutils} programs are similar to the types seen in this study. 
$\realcodeTypeBmsCoverageRatio$\% of the data types in \texttt{uutils} programs also exist in translations by our users. 
The most frequent Rust types in \texttt{uutils} programs are string-related types such as \code{String} ($\realcodeTypeStrRatioString$\%), \code{&str} ($\realcodeTypeStrRatiostr$\%), \code{OsString} ($\realcodeTypeStrRatioOsString$\%), and \code{OsStr} ($\realcodeTypeStrRatioOsStr$\%). One difference is that \code{OsString} and \code{OsStr} types do not exist in translations by our users.} 
\todoadd{These types bridge the gap between platform-native strings and Rust \code{String}s and are useful for \texttt{uutils/coreutils} as it aims to be cross-platform. Our user study only specified the requirements that the Rust code should work on Linux.}

\noindent\todoadd{\textbf{Library API Calls.} Recall that in our user study, users translate many C library API calls using various strategies listed in Fig.~\ref{fig:fig_api_mapping}. We also investigate how library API calls in GNU C coreutils are expressed in the \texttt{uutils} programs. 
Due to a large number of library API calls ($561$ in total) and large program dissimilarity, we focus on the top $10$ frequently used APIs that covered $53$\% of all the API calls. 
We manually look into at least 5 call sites per API and check Rust code fragments that implement corresponding functionalities. Most API calls in C can be mapped to safe Rust API calls and expressions.
Third-party crates are used for some APIs. One example is the \code{quote} C API that handles special characters. The Rust program implements the same functionality using \code{quote} API provided by the Rust crate \code{os_display}. We find no API calls handled using the API emulation strategy, however.}

\noindent\todoadd{\textbf{Dealing with Aliasing.} Recall that users in our study use strategies including (A) \emph{reference elision} and (B) \emph{cloning} to deal with aliasing (Section~\ref{sec:aliasing}). For the \texttt{uutils} programs, we find C code locations involving aliasing pointers and then check how a similar functionality is implemented in Rust. Due to large dissimilarities from the C source, we manually check around $20$ places in C and find $4$ with clear matching code fragments in Rust. Those code fragments are all examples of \emph{reference elision}, and one worth mentioning is the mapping of a linked list in C to a \code{Vec} in Rust in the \texttt{csplit} program. Such type lifting eliminates code blocks with aliasing pointers that manipulate linked lists.
We also find many instances of object cloning in Rust \texttt{uutils} to satisfy ownership constraints when calling functions\tododel{of a callee function at various call sites}. But none are specifically used for handling aliased references as was done in our user study.}

\noindent\todoadd{\textbf{Functional Equivalence.} In Section \ref{sec:testdiff}, we reported that user translations are not fully equivalent to the C source and multiple differences are exposed by differential fuzzing.
The phenomenon is observed in the \texttt{uutils} project, which currently passes around 80\% of the GNU test suite.
We find that for 5 of the 6 \texttt{uutils} programs we investigated, there are known correctness gaps with GNU Coreutils when using its test results, as reported in the \texttt{uutils} repository\footnote{\todoadd{According to their \href{https://github.com/uutils/coreutils-tracking/blob/6cd735e861ee67773ad2dee41157d8abc6796db7/gnu-full-result.json}{\texttt{gnu-full-result.json}} of commit a0d258d.}}.
One program, \texttt{test}, seems to be close to functional equivalence since it passes the full GNU test suite.
We further compare the Rust version of \texttt{test} to the GNU C version using differential fuzzing in a setup similar to Section \ref{sec:testdiff}.
Among $300$ fuzzer-generated tests sampled, we find $\realcodeEquivPertFail$\% test cases exhibit different behaviors.
With a deeper investigation, we find that at least $\realcodeEquivPertLogicalAndErrmsg$\% of the tests reveal non-trivial logical differences. 
More specifically, C and Rust programs have different results (i.e., return code) on $\realcodeEquivPertLogical$\% tests and semantically different error messages on $\realcodeEquivPertErrmsg$\% tests.
We observe no I/O encoding errors or runtime safety aborts.
We also find that the maintainers are aware of several semantic differences that need to be fixed according to comments in their additional test file.}

\noindent\todoadd{\textbf{Use of Unsafe Rust.} 
Among the $6$ Rust programs, only one program (\texttt{test}) has one line of unsafe Rust to call \code{libc::isatty}. This C library call checks if the open file descriptor is a terminal. It can perhaps be replaced with safe Rust APIs in third-party crates, such as \code{atty} or \code{termion}.
}

\begin{tcolorbox}[breakable,width=\linewidth,boxrule=1pt,top=0pt, bottom=0pt, left=1pt,right=1pt]
\todoadd{Most of the findings in our user study are also applicable to a Rust project mirroring GNU Coreutils written by experts.}
\end{tcolorbox}

\section{Discussion and Takeaways}
\label{sec:discussion}

Our analysis of user-provided translations highlights several points where automatic translation strategies deviate from those taken by human users. We reflect on why there is such a gap and summarize takeaways for automatic translation.

\noindent\textbf{Separate Policy from Mechanism.}
Our observation is that there are many choices to be considered when translating a C program to Rust, i.e., there is no one-size-fits-all strategy to take. The resulting translations, while satisfying memory safety, can have varying levels of functional correctness, performance, and grace in handling runtime safety exceptions. The balance between them 
is a matter of {\em policy}. 
\todoadd{
For example, the level of functional equivalence to achieve is one crucial policy decision.
Full functional equivalence can require significant effort and may even go against the purpose of code migration~\cite{verhoef2000realities}. 
To characterize what behaviors must be kept and what behaviors can change in the translation, we may need more thorough unit tests or other forms of specifications.
Another example of an important policy decision is about the data types and APIs to use, which can lead to multiple ways of translating the same program\tododel{. Different data types, APIs, or third-party dependencies can be used to implement the same functionality} with different trade-offs in performance, memory overhead, compatibility, and so on. 
In summary, users may want explicit control over policy decisions even when using automatic tools.}

\noindent\textbf{Improving LLM-based Search for Translations.}
Once policy decisions are clear, automatic mechanisms can enable search for translations. Our observation is that there are several immediate and open problems that, if addressed, can make C to Rust code automatic translator much more usable.

\noindent\emph{(1) Modeling of Data Types and APIs.} 
So far, few of the previous work explicitly models data types and APIs in the Rust standard library, such as the ones summarized in Fig.~\ref{fig:typemap}.
We believe it is essential to overcome the language differences between C and Rust and move away from unsafe code.\\
\noindent\emph{(2) Mergeable Decomposition.} Finding the right way to decompose programs into smaller components is challenging but appears necessary for LLMs with limited context windows. 
At the same time, decomposition need not be one-shot. It is useful to explore if it is possible to incrementally transform identified components, such that a partial Rust translation can be used alongside the untranslated part of C. 
\\
\noindent\todoadd{\emph{(3) The Last Mile Problem.}
Some users reported in their (optional) qualitative feedback about their experience. $10$ users reported that debugging was tedious in locating the root cause of certain semantic differences.
$6$ users mentioned that fixing some semantic differences involving library calls (e.g., \code{regex}) is not easy. 
\todoaddsec{$14$ users mentioned that LLMs are ineffective in directly generating correct long code due to issues like type inconsistency across functions and multiple inter-related errors.} 
4 participants who used LLMs for assistance reported that they often needed to restart with clean context to get better output.}

\noindent\todoadd{\textbf{Promising Static or Dynamic Analysis.} Based on our findings, we foresee $3$  program analyses as immediately useful:}

\noindent\todoadd{\emph{(1) Refactoring Global Variables Through Lifetime and Def-Use Analysis.} Use of mutable globals differentiates C and safe Rust programs. 
As observed in our user study, many globals are accessed within a thread and can be moved to the heap or stack, possibly in grouped \code{struct}-typed objects for efficiency. Precise analysis of lifetime of globals is a promising next step.\\
\noindent\emph{(2) Lifting Data Types Semantically Before Lifting References.} Data type analysis directly impacts how references are lifted to Rust. Code blocks and pointers in C can often be elided after type lifting, which is observed from both translations by our users and in our extensibility analysis on public code(e.g., Linked list pointer manipulations to \code{Vec} API calls). LLMs can be useful for pattern recognition tasks, according to qualitative feedback received from our users, and it would be promising to use them for type mapping suggestions of C objects to Rust.\\
\noindent\emph{(3) Lifting Unions.} Automatic analysis to (a) tell apart different use cases of C unions and (b) associate tag values with valid fields for variant records can help convert unions to enums.}

\noindent\textbf{Threats to Validity.}
Our user study is with a relatively small number of users and on relatively small programs. These present limits are an artifact of time constraints and the inherent difficulty of the task at hand for users who have limited prior experience with Rust.
\todoadd{We point out that Rust has a relatively new developer community. A survey by JetBrains in 2023 involving over $26k$ developers worldwide reports the majority of Rust developers (56\%) have less than $6$ months of experience~\cite{rustsurvey}. 
To partially assess which of our findings extend to code written by more experienced developers, we reported on our post-hoc analysis on more mature real-world Rust programs mirroring GNU C coreutils in Section \ref{sec:experts}. Most of the key findings we highlight extend beyond our users.}

\section{Related Work}
\label{sec:related}

In practice, many motivations for code migration to new
languages exist, such as legacy code modernization\tododel{for maintenance}, compatibility improvement with newer platforms, N-versioning for fault tolerance, and so on. Memory safety is a unique driver for C to Rust translation. 
\tododel{For example, about $11.5\%$ of the Mozilla Firefox browser codebase, which is originally C/C++, has been ported to Rust with safety as a key consideration~\cite{Mozilla}.}%
\todoadd{For example, Mozilla Firefox is actively migrating components written in C/C++ to Rust with memory safety as a key consideration~\cite{oxidation}.}\tododel{The latest statistics show that more than 10\% of the Firefox codebase is in Rust~\cite{ffstats}.} Towards memory safety as a goal, a long line of memory defenses has been investigated. They achieve different trade-offs between performance, compatibility, and security.

\noindent\textbf{Memory Safety in C/C++.}
SoftBound~\cite{softbound} and CETS~\cite{cets} are compiler-based solutions that provide full memory safety in C. \tododel{Spatial safety can be achieved by associating bounds information with pointers and checking them at runtime. Temporal safety can be achieved using lock-and-key abstractions.} The performance overheads of enforcing full memory safety \todoadd{in software-based approaches} are often reported to be higher than $50\%$.
In recent years, specialized hardware features have emerged to accelerate both spatial~\cite{cheri,capstone,PAC,MPK} and temporal checks~\cite{ARMMTE,capstone}, but overheads below $10-20\%$ for full safety appear elusive. Runtime defenses for spatial safety can be acceptably low~\cite{sticktags},
but complete temporal safety still bears a bulk of the runtime overheads. 
\todoaddsec{Languages like Rust can help developers explicitly manage object lifetimes in a way
that eliminates temporal memory management mistakes.  In our user study, we see that users make heavy use of zero-cost abstractions for temporal safety.}
Owing to the costs of full memory safety, partial safety defenses have been investigated and have a rich history~\cite{sokwar}. Prominent among these are CFI~\cite{cfi2014}, ASLR~\cite{aslr}, stack canaries~\cite{canary}, guard pages~\cite{Chiueh2001RADAC}, and DEP~\cite{dep} which have found wide deployments.
These defenses have good performance characteristics but do not rule out all memory safety errors. For example, exploits that bypass these deployed defenses without violating control-flow properties~\cite{dop, heartbleed} are known~\cite{chrome-bypass}.

\todoaddsec{Another approach is based on proactive discovery of bugs and subsequent automatic
repair at the source C code level. Greybox fuzzing~\cite{aflplusplus,ossfuzz}, symbolic execution~\cite{klee,symcc}, and their combination~\cite{hybrid} for finding security vulnerabilities is an active area of research. Automatic localization of buggy code~\cite{faultloc,senx} and generation of suggestions for fixes are being actively explored~\cite{gao2022programrepair,song2024provenfix}. This find-and-fix approach offers a continual process to improve software quality and reduce patching effort once flaws are discovered.}

\noindent\textbf{Memory Safe Dialects for C.}
Writing code that is free of memory safety bugs is a desirable goal. Several works focus on finding language subsets or extensions of C/C++ that are easier to statically analyze and dynamically retrofit safety checks than untamed C.  CCured explored spatial safety via ``fat'' pointers and relies on garbage collection for temporal memory safety~\cite{ccured}.  
Xu et. al achieved temporal memory safety using a global capability store instead of a garbage collector~\cite{xu2004efficient}. Cyclone has similar spatial
safety mechanisms as CCured but uses memory regions
for temporal safety~\cite{cyclone}. Flow-sensitive type qualifier analysis can leverage user-provided type annotations for static analysis~\cite{foster2002flow}.
\tododel{The goal of those language designs is often to retain C's low-level abstractions while offering safety for a subset of C, which limits their compatibility.}
More recently, work on the Checked C language~\cite{checkedc} aims to enable mixed legacy C pointers with safe ones for incremental migration, by allowing parts of the code to be type-annotated and proven safe. However, it does not provide full memory safety. Efforts to dispatch more spatial safety statically at compile-time are underway, which can further reduce costs of spatial safety~\cite{deputy,corecheckedc,3c}. 
There are also efforts to introduce temporal safety into Checked C with runtime overheads of about $30\%$ or more~\cite{zhouoopsla23}. 
Overall, while designing safe C dialects continues to be a promising endeavor, low-overhead designs that eliminate all memory safety bugs have yet to be found.

\noindent\textbf{Translating C to Rust.}
Safe Rust abstractions force developers to move away from raw C pointers. The abstractions offered by safe Rust---lifetime~\cite{lifetime}, ownership~\cite{ownershipilya}, and the AXM principle~\cite{lifetime}---force a significant departure from untamed C or C dialects in how programmers write code. 
\todoadd{In the previous sections, we have compared various previous work~\cite{emre2021translating,zhang2023ownership,eniser2024towards,yang2024vert}, and scalable solution is still an open challenge.}

\section{Conclusion}
\label{sec:conclusion}
We have presented the lessons learned from a user study on how users can translate C programs to safe Rust, with good performance and security gains. Our analysis reveals that they share a high-level approach and common specialized strategies to overcome the challenges that encumber automatic tools. Zero-cost abstractions are ubiquitously used, significantly reducing the costs of runtime temporal safety checks, which highlights why Rust offers a promising road ahead. 
\tododel{We hope to inform future work on automatic C to Rust translators.}

\section*{Acknowledgment}
\todoaddcam{We thank all the participants in this user study.} We also thank the anonymous reviewers and our shepherd for giving us valuable feedback on an earlier draft of this paper. This research is supported in part by the research funds of the Crystal Centre at the National University of Singapore and the Cisco University Research Program Fund, a corporate advised fund of Silicon Valley Community Foundation.

\bibliographystyle{IEEEtranS}
\bibliography{IEEEabrv,paper.bib}

\appendix

\subsection{Case Studies with Memory Safety Vulnerabilities}
\label{sec:apdxvuln}

\noindent\emph{\textbf{Case Studies}}. We took a closer look at the translations of \texttt{shoco} and \texttt{urlparser}, which are the two C programs with spatial and temporal memory errors.
We investigate if the memory errors are eliminated or detected at runtime in the Rust translation.

\noindent\emph{Case Study 1: \texttt{shoco} library}. 
\noindent The C string data compression library \texttt{shoco} has one spatial memory vulnerability (CVE-2017-11367)\footnote{\url{https://nvd.nist.gov/vuln/detail/CVE-2017-11367}}. The global array \code{packs} will be overread when the byte to be decompressed is malformed.
In this case, the return value by the header decoding function will be larger than the length of \code{packs} array. When the return value is used to index \code{packs}, the global buffer overflow happens, as highlighted in line \todo{5} in \todoaddcam{Fig~\ref{fig:case_study} (a)}.

This spatial memory issue is detected at runtime in all Rust translations from our users. 
Since array \code{packs} is read-only, most translations lift it to a constant global array with type \code{[Pack; 3]}.
Rust's spatial bounds checking on arrays ensures any out-of-bound array access will not happen. The Rust compiler inserts runtime checks before the index access \code{PACKS[mark]}. This is because it cannot statically infer the validity of this access since the value of \code{mark} is determined by the input. 
Fig~\ref{fig:case_study} (b) shows one translation, where the binary will panic if the buffer overflow is about to happen.

\todoadd{It is worth noting that existing \texttt{c2rust}-based tools (including \laertes and \crown) are also able to eliminate this vulnerability. They perform the same syntactic transformation in their first stage to convert the global array declaration in C to an equivalent declaration in Rust. Such array declarations are bounds-checked on access. Since their translation of this global array and the access are similar, we show one example of the translated code snippet by their tools relevant to the vulnerability in Fig. \ref{fig:case_study} (c).}

\begin{figure}[th]
    \centering
\begin{minipage}{\linewidth}
(a) The source code related to a spatial memory issue
\begin{lstlisting}[language=C, style=boxed]
static const Pack packs[PACK_COUNT] = { ... };

size_t shoco_decompress(...) {
    while (in < in_end) {
        if (mark < 0) { ... }
        else {
            if (o + packs[mark]...) // Global buffer overflow!
        }
    }
}
\end{lstlisting}
\end{minipage}
\begin{minipage}{\linewidth}
(b) Safe Rust translation of the above C code by our user
\begin{lstlisting}[language=Rust, style=boxed]
const PACKS: [Pack; PACK_COUNT] = [ ... ];

fn shoco_decompress(...) {
    while in_index < in_end {
        if mark < 0 { ... }
        else {
            if o_index + PACKS[mark as usize]... // BoF caught by runtime check
        }
    }
}
\end{lstlisting} 
\end{minipage}
\begin{minipage}{\linewidth}
(c) Rust translation by compiler-based tools (\laertes/\crown)
\begin{lstlisting}[language=Rust, style=customrust]
static mut packs: [Pack; 3] = ...;

pub unsafe extern "C" fn shoco_decompress(...) {
   ...
   while in_0 < in_end {
      mark = ...;
      if ... {} 
      else {
            if o.offset(packs[mark as usize].bytes_unpacked as size) > out_end ... // BoF caught by runtime check
      }
   }
}
\end{lstlisting} 
\end{minipage}
    \caption{Case Study 1: The spatial memory issue in the \texttt{shoco} library is detected in all Rust translations.}
    \label{fig:case_study}
\end{figure}

\begin{figure}[th]
    \centering
\begin{minipage}{\linewidth}
(a) Code example related to the temporal memory issue
\begin{lstlisting}[language=C, style=boxed]
// urlparser.c
url_data_t *url_parse (char *url) {
  url_data_t *data = malloc(sizeof(url_data_t));
  data->href = url; // Store the pointer
}
void url_data_inspect (url_data_t *data) {
  printf("#url =>\n");
  printf("    .href: \"%s\"\n", data->href); // use the pointer
}

// poc.c
int main() {
    // our_url points to a URL string stored on the heap
    url_data_t *parsed = url_parse(our_url);
    assert(parsed);
    free(our_url); 
    url_data_inspect(parsed); // Use-after-free here!
}
\end{lstlisting}
\end{minipage}
\begin{minipage}{\linewidth}
(b) One corresponding Rust translation of the above C code
\begin{lstlisting}[language=Rust, style=boxed]
fn url_parse(url: &str) -> Option<UrlData> {
    // data is a default UrlData instance
    data.href = Some(url.to_string()); 
    // href has Option<String> type
    // ...
}
fn url_data_inspect(data: &UrlData) {
    println!("#url =>");
    println!("    .href: {:?}", data.href); // Owned String. No temporal issue.
}
\end{lstlisting} 
\end{minipage}

    \caption{Case Study 2: The temporal memory issue in the \texttt{urlparser} library is eliminated in all Rust translations.}
    \label{fig:case_study2}
\end{figure}

\noindent\emph{Case Study 2: \texttt{urlparser} library}.
\noindent The \texttt{urlparser} C library (commit \href{https://github.com/jwerle/url.h/tree/a65623ad107be19ca4efb5a36379f3440eb48091}{\texttt{a65623ad}}) has multiple memory-related vulnerabilities, including a spatial memory vulnerability that causes information leaks and a temporal memory vulnerability that can potentially cause programs using this library to crash. Both vulnerabilities have been fixed in their latest version.

\noindent\textbf{Temporal memory errors.} 
This library aliases the input URL string pointer and stores it as a field in its custom structure when parsing this string. In addition to the parsing API, it offers APIs that read the parsing result, such as the \code{inspect} API. 
This is problematic because this library does not own the memory referred to by this aliased pointer. 
Use-after-free can happen if a program using this library calls the parse API 
providing a string that does not live long enough before calling other APIs, such as \code{inspect}.
Fig.~\ref{fig:case_study2} (a) shows the related source code and a small example program that triggers use-after-free. This issue has been fixed in commit \texttt{752635e}.

This temporal issue is eliminated in all Rust translations because of the temporal safety statically enforced by borrowing rules and the \code{String} type. 
Any code pattern that can result in double-free or use-after-free cannot be compiled. 
To satisfy borrow rules, user translations copy the input string so that the struct representing the parsing result can hold an owning reference to the string, as shown in Fig.~\ref{fig:case_study2} (b). This copy is automatically deallocated when the lifetime of the parsing result (i.e., \code{UrlData}) ends. 

\noindent\textbf{Spatial memory errors.}
This library can overread the input string and print extra memory value if the input URL is malformed. 
The root reason is that it directly increments the URL pointer to skip the scheme separator \code{://} without checking if the separator exists in the URL. When it does not exist, the pointer is incremented to be outside of the input string buffer.
This bug is caught at runtime by all Rust translations. This pointer is lifted to \code{&str} or \code{&String} types, which enforce spatial bounds checking on access.
Several other heap buffer overflow issues exist as well, all due to saving strings to allocated memory regions with insufficient space, i.e., having an allocated size smaller than the string length. Such spatial issues do not exist in safe Rust translations by users due to automated memory management. 
\subsection{Examples of Logical Translation Errors by Users}\label{sec:applogic}

Here we describe several logical translation errors we observed in translations of \texttt{fmt} program, along with a possible fix for each.

\noindent\textit{1) Logical Errors that Require More Thorough Testing}

The full program of the translation (Version A) we have seen before (Fig. \ref{fig:exp_lifting_rs1}), although passing all tests with 85\% code coverage, has at least two more semantic discrepancies from the source program. Surprisingly, the bugs are in code lines covered by passing tests.
We have two failing tests demonstrating each of the translation bugs in Fig. \ref{fig:exp_diff0}.
  One bug is revealed when there are multi-byte characters rather than just ASCII characters in the input. Another bug is revealed when there are odd numbers rather than even numbers of padding spaces. 
  The unit tests written by the user miss those cases.

The first discrepancy is due to calling the wrong API when computing the display width of the characters. The translation used \code{char::len_utf8} which computed the size in bytes of a Unicode character instead of display width. 
          For the frequent ASCII characters, the size in bytes happens to be 1, which coincides with their display size. However, the failing input "\code{z\\u00df\\u6c34\\U0001d10b}" is an example where the size in bytes is different from the display width. 
          It is not revealed by the passing unit tests which only include ASCII characters.
          The fix is to use the \code{c.width} method, as shown in Fig.~\ref{fig:exp_diff0}.

The second discrepancy is due to an off-by-one error in the computation of padding spaces, making the Rust translation correct only when the number of spaces is even. 
          The Rust program uses a slightly different computation to split the padding spaces into leading and trailing ones, compared to the C program.
          While the C program behaves like a round-up division, the Rust program takes the floor division.
          The passing test does not reveal this difference.
          The correct fix is shown in Fig. \ref{fig:exp_diff0}.

\begin{figure}[ht]
\centering
\begin{minipage}{\linewidth}
Passing Test (Translation A)
\begin{lstlisting}[basicstyle=\ttfamily\footnotesize]
Executed command: ./fmt -c -w 10
Input: "Center"
C Output: "  Center" // 2 leading padding spaces
Rust Output: "  Center" // same as C
\end{lstlisting}
Failing Test 1
\begin{lstlisting}[basicstyle=\ttfamily\footnotesize]
Executed command: ./fmt -c -w 10
Input: "z\u00df\u6c34\U0001d10b"
C Output and Rust Output have different display width
\end{lstlisting}
Failing Test 2
\begin{lstlisting}[basicstyle=\ttfamily\footnotesize]
Executed command: ./fmt -c -w 10
Input: "Center*"
C Output: "  Center*" // 2 leading padding spaces
Rust Output: " Center*" // off-by-one error!
\end{lstlisting}
The bugfix (Translation A)
\begin{lstlisting}[
  basicstyle=\ttfamily\footnotesize,
  frame=single,
  commentstyle=\color{gray},
  keywordstyle=\color{blue},
  lineskip=-0.4ex,
  language=Rust,
  style=diff
]
...
+++ use unicode_width::UnicodeWidthChar;++
if bytes_read == 0 { break; }
let len: usize = p1.trim().chars().map(
   |c| if c == '\t' { ' ' } else { c }
--- ).map(char::len_utf8).sum();--
+++ ).map(|c| c.width().unwrap_or(1)).sum();++
let padding = 
--- (config.goal_length - len) / 2;--
+++ (config.goal_length - len + 1) / 2;++
...
\end{lstlisting}
\end{minipage}
    \caption{The Translation Version A wrongly computes the display width. Part (a): 1 passing and 2 failing tests when aligning the input centrally in a 10-byte line (executed command: \code{./fmt -c -w 10}). Part (b): Bugfix for two semantic discrepancies.}
    \label{fig:exp_diff0}
\end{figure}

\noindent\textit{2) Semantic Differences that are Not Easy to Fix}

There is another difficult-to-fix semantic discrepancy that exists in all user translations for the \texttt{fmt} program. They all omit the functionality of replacing invalid Unicode characters with \code{?}. With a deeper investigation, we find that this is not easy to fix due to various differences in data types and APIs across the two languages.

\begin{figure}[th]
    \centering
\begin{minipage}{\linewidth}
\begin{lstlisting}[
  basicstyle=\ttfamily\scriptsize,
  commentstyle=\color{gray},
  keywordstyle=\color{blue},
  backgroundcolor=\color{gray!10},
  lineskip=-0.4ex,
  language=C,
  style=highlightc
]
static void
center_stream(FILE *stream, const char *name)
{
    char *p1, *p2;
    wchar_t wc;
    size_t len;	/* Display width of the line. */
    int w1;	/* Display width of one character. */
    int w2;	/* Length in bytes of one character. */
    
    while ((p1 = get_line(stream)) != NULL) {
        len = 0;
          for (p2 = p1; *p2 != '\0'; p2 += w2) {
              // ... omitted
\end{lstlisting}
\vspace{-0.5\baselineskip}
\begin{tcolorbox}[mylistingstyle]
\begin{lstlisting}[
  basicstyle=\ttfamily\scriptsize\bfseries,
  commentstyle=\color{gray},
  keywordstyle=\color{blue},
  backgroundcolor=\color{gray!1},
  lineskip=-0.4ex,
  language=C,
  style=highlightc
]
              if ((w2 = mbtowc(&wc,p2,MB_CUR_MAX))==-1) {
                  (void)mbtowc(NULL,NULL,MB_CUR_MAX);
                  *p2 = '?';
                  w2 = 1;
                  w1 = 1;
              } else if ((w1 = wcwidth(wc)) == -1)
                  w1 = 1;
\end{lstlisting}
\end{tcolorbox}
\vspace{-\baselineskip}
\begin{lstlisting}[
  basicstyle=\ttfamily\scriptsize,
  commentstyle=\color{gray},
  keywordstyle=\color{blue},
  backgroundcolor=\color{gray!10},
  lineskip=-0.4ex,
  language=C,
  style=highlightc
]
            // ... omitted
        }
        // ... print whitespace padding
        puts(p1);
    }
    // ...
}
\end{lstlisting}
\end{minipage}
    \caption{A highlighted block of code of the C example program (\texttt{fmt}) that deals with invalid characters (previously omitted for simplicity). }
    \label{fig:exp_lifting_add}
\end{figure}

\begin{figure}[th]
    \centering
\begin{minipage}{\linewidth}
\begin{lstlisting}[language=Rust, style=boxed]
use unicode_width::UnicodeWidthChar;
fn center_stream<R: BufRead>(mut stream: R, _name: &str, config: &Config) {
    let mut p1: Vec<u8> = vec![];
    while let Ok(bytes) = stream.read_until(b'\n', &mut p1) {
        if bytes == 0 { break; }

        let mut res = String::new();
        let mut idx = 0;
        loop {
            let remain: &[u8] = &p1[idx..];
            match remain.utf8_chunks().next() {
                Some(s) => {
                    let valid = s.valid();
                    if valid.is_empty() {
                        res.push_str("?");
                        idx += 1;
                    } else {
                        res.push_str(valid);
                        idx += valid.len();
                    }
                },
                None => {
                    break;
                }
            }
        }
        let len = ...; 
        // ... remaining lines omitted
    }
}
\end{lstlisting}
\end{minipage}
    \caption{The fixed Translation A where the data type of \code{p1} is refined from \code{String} to \code{Vec<u8>}. The code to parse the Unicode character is updated accordingly. }
        \label{fig:exp_lifting_fix}
\end{figure}

Fig. \ref{fig:exp_lifting_add} highlighted the branch of the C code to handle this functionality, which was previously skipped in Fig.~\ref{fig:exp_lifting_c} for simplicity.
All users assigned to this program omit this functionality in their translation.

With closer examination, it turns out that this behavior is not easily implementable when using the \code{String} type to represent \code{p1}.
In Rust, characters stored in a \code{String} data type are limited to Unicode characters rather than arbitrary raw bytes.
When using the Rust API \code{read_line} to read a \code{String} from the stream, it returns either a \code{String} instance or a \code{Err} result if there are invalid characters.
When returning an \code{Err}, there is no API to tell where the invalid characters are.
With a more careful search, we find an API \code{from_utf8_lossy} on the \code{String} type in Rust that is closer to what we want. This API can create a \code{String} from a buffer with each point of decoding error marked with \code{?}. 
However, we soon realize that this API also does not bridge the semantic gap. The C code not only outputs \code{?} at the position of invalid characters, but also outputs the same number of \code{?} as the number of invalid bytes.
\todoaddcam{The \code{from_utf8_lossy} API is not behaving the same way as the corresponding C code---it replaces each chunk of invalid bytes as one \code{?}. For example, for the invalid byte sequence 
\code{"\\x7a\\xc3\\x9f\\xe6\\xb0\\xc3\\xf0\\x9d\\x84\\x8b"}, the Rust String API \code{from_utf8_lossy} outputs a string with two \code{"?"} while the C code outputs three \code{"?"}.}
In summary, we cannot find APIs on \code{String} to preserve the same semantics.

As a result, we can consider refining the choice of Rust data types for \code{p1}. 
The type of \code{p1} should carry sufficient information and support APIs to identify (1) the locations of invalid characters, and (2) the number of invalid bytes at each location. 

Many Rust data types can represent a string but not all of them can meet our requirements. 
One possible choice is \code{std::ffi::OsString}, which is a platform-native string type that may hold invalid UTF-8 characters.
\code{Vec<u8>} is another choice that can store invalid Unicode characters and supports per-byte control.
Here we use \code{Vec<u8>} as an example.

Now the question is about how to use \code{Vec<u8>} to achieve the intended program behavior, i.e., to parse Unicode characters and replace invalid ones with the correct number of question marks. 
    One option is to use \code{std::str::Utf8Chunks} structure from the Rust standard library. This structure allows iteratively check \code{u8} bytes referred by an immutable slice and convert them to Unicode characters if valid.
Fig.~\ref{fig:exp_lifting_fix} shows the fixed version of Translation A with the correct Unicode character parsing logic when using \code{Vec<u8>} type for \code{p1}. 
Because of this new choice of type, the original code to read lines and parse characters needs to be updated accordingly.
We update multiple lines in the method calls on \code{stream} to read the input and iteratively save the line-separated input into a \code{Vec<u8>}.
The eventually fixed translation (shown in Fig. \ref{fig:exp_lifting_fix}) can pass the tests (behave the same as the C program) on the aforementioned invalid byte sequence.

\subsection{\todoadd{End-to-End Performance Results of User Translations}}
\label{sec:e2eperf}

\todoadd{Besides the performance results explained in Sec. 
\ref{sec:perf}, we also measure the end-to-end performance on tests using hyperfine~\cite{hyperfine}, which considers the load/start-up time of programs, in addition to the execution of the main functionality. The average overhead is around 13\% across the eight Rust translations. Details of the performance measure for each program are shown in Fig. \ref{fig:hyperperf}. This measure might not reflect the performance we care about since initialization costs are usually amortized for long-running programs and libraries. }

\begin{figure}[ht]
  \centering
  \begin{subfigure}[b]{0.49\linewidth}
    \includegraphics[width=\linewidth]{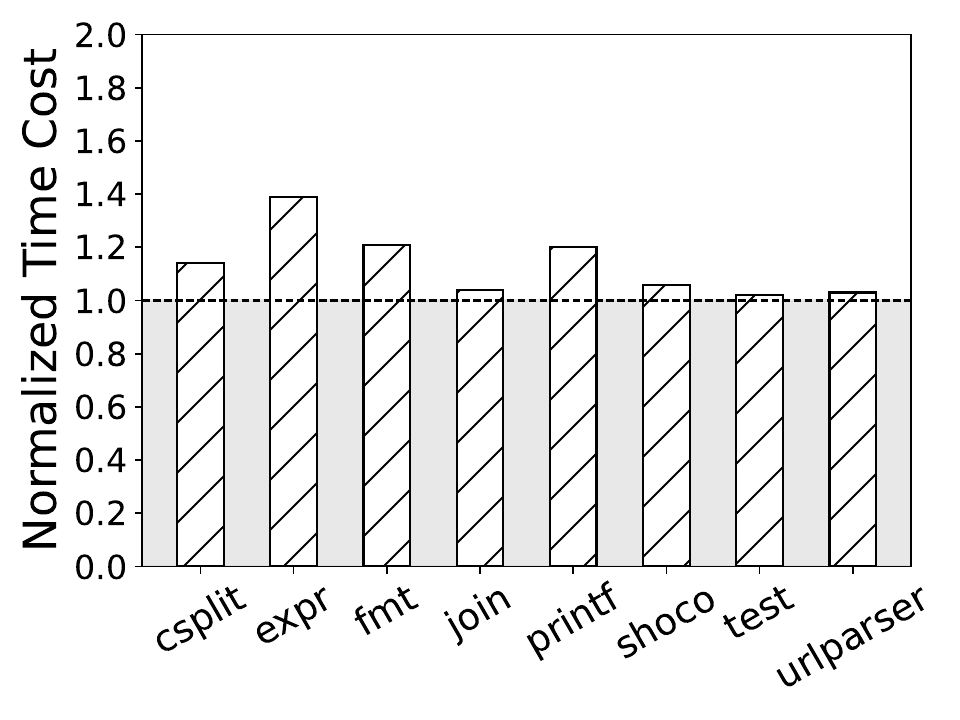}
    \caption{Compiled with \code{-O2}}
    \label{fig:hyperperfcmp2}
  \end{subfigure}
  \hfill
  \begin{subfigure}[b]{0.49\linewidth}
    \includegraphics[width=\linewidth]{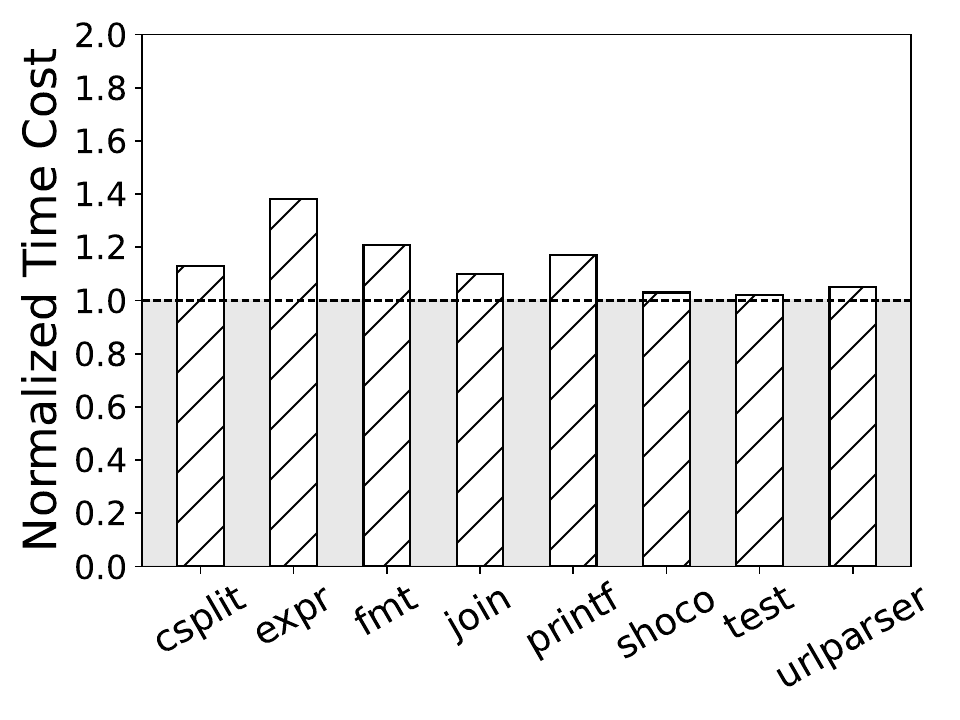}
    \caption{Compiled with \code{-O3}}
    \label{fig:hyperperfcmp3}
  \end{subfigure}
  \caption{\todoadd{End-to-end running time measured by hyperfine~\cite{hyperfine} of the most similar Rust translation compared to the original C at different optimization levels. The time for the baseline C (dashed line) is normalized to 1.}}
  \label{fig:hyperperf}
\end{figure}

\ifdefined\confver

\else

\subsection{\todoadd{Details of Code Patterns involve Globals and Unions}}
\label{sec:apdxpattern}

\noindent\todoadd{\textbf{Translating Mutable Globals.} There are four strategies observed from the user translations on the translation of global, as previously explained in Sec. \ref{sec:safety2}. Here, we show a code example for each strategy. }

\begin{figure}[th]
    \centering\footnotesize
\begin{minipage}{\linewidth}
\texttt{Translating Mutable Globals (Strategy A)}
\begin{lstlisting}[language=C, style=customc]
// printf.c
char **gargv;
static int getchr(void)
{
    if (!*gargv)
        return((int)'\0');
    return((int)**gargv++);
}
...
\end{lstlisting}
\end{minipage}
\begin{minipage}{\linewidth}
\begin{lstlisting}[language=Rust, style=boxed]
// printf.rs
fn getchr(argv: &Vec<String>, idx: &mut usize) -> char {
    if *idx >= argv.len() {
        return '\0';
    }
    let c: char = argv[*idx]...; // access argv
    *idx += 1; // change idx (equivalent to incrementing gargv)
    return c;
}
int main() {
    let mut argv: Vec<String> = env::args().collect();
    let mut idx: usize = 1; // globals moved to stack
    // ...
    let c: char = getchr(&argv, &mut idx); // passing references
}
...
\end{lstlisting}
\end{minipage}
    \caption{(Top) A part of the \texttt{printf} C program with a mutable \code{gargv}. (Bottom) A Rust translation using strategy A: Moving globals to locals.}
    \label{fig:globalparam}
\end{figure}

\begin{figure}[th]
    \centering\footnotesize
\begin{minipage}{\linewidth}
\texttt{Translating Mutable Globals (Strategy A + Structs)}
\begin{lstlisting}[language=C, style=customc]
// fmt.c
static int centerP = 0;	
static size_t x;
// ... 15 globals in total
static void new_paragraph(size_t indent)
{
    // ... modifying 4 globals 
    x = indent;
}
...
\end{lstlisting}
\end{minipage}
\begin{minipage}{\linewidth}
\begin{lstlisting}[language=Rust, style=boxed]
// fmt.rs
pub struct TextInfo {
    center: bool, // globals become struct fields
    x: usize,
    // ... other semantically related fields
}
impl TextInfo {
    pub fn new_paragraph(&mut self, indent: usize) {
        // ... modifying multiple fields in &mut self
        self.x = indent;
        ...
    }
}
...
\end{lstlisting}
\end{minipage}
    \caption{(Top) A part of the \texttt{fmt} C program with mutable globals. (Bottom) A Rust translation using strategy A and group globals into structs.}
    \label{fig:globalgroup}
\end{figure}

\todoadd{\emph{(A) From Globals to Locals.} Fig. \ref{fig:globalparam} shows an example where globals are moved to stack. The top of Fig. \ref{fig:globalparam} shows a piece of code from the \texttt{printf} C program that declared a mutable global variable named \code{gargv}. This variable is used as a cursor that shifts around to access arguments (\code{argv}) passed to the \code{main} function. A Rust translation by our user simply declares what is equivalent to such a cursor on the stack (\code{argv} and \code{idx}). The Rust code passes them by reference to the function that needs to access them (i.e., \code{getchr}).}

\todoadd{When there are many global variables, globals that are used together with similar lifetimes can be grouped into structs. Fig. \ref{fig:globalgroup} shows a piece of code from the \texttt{fmt} program in our benchmark that uses more than ten mutable global variables for sharing states between functions. A user translation (bottom of Fig. \ref{fig:globalgroup}) groups the relevant global variables that are accessed together into a struct named \code{TextInfo}. Operations on those global variables (such as the \code{new_paragraph} function) are turned into methods implemented on the struct \code{TextInfo}.}

\begin{figure}[th]
    \centering\footnotesize
\begin{minipage}{\linewidth}
\texttt{Translating Mutable Globals (Strategy B)}
\begin{lstlisting}[language=C, style=customc]
// csplit.c
long	 nfiles;
void cleanup(void)
{
    char fnbuf[PATH_MAX];
    for (i = 0; i < nfiles; i++) {
        snprintf(fnbuf, ...) // details omitted
        unlink(fnbuf);
    }
}
...
\end{lstlisting}
\end{minipage}
\begin{minipage}{\linewidth}
\begin{lstlisting}[language=Rust, style=boxed]
// csplit.rs
static FILES: Lazy<Arc<Mutex<Vec<String>>>> = Lazy::new(|| ...);
// ref-counted with lock
fn cleanup() -> io::Result<()> {
    let mut files = FILES.lock().unwrap();
    for file in files.iter() {
        fs::remove_file(file)?;
    }
    files.clear();
    Ok(())
}
...
\end{lstlisting}
\end{minipage}
    \caption{(Top) A part of the \texttt{csplit} C program with mutable globals. (Bottom) A Rust translation using strategy B: dynamic references.}
    \label{fig:globaldyn}
\end{figure}

\begin{figure}[th]
    \centering\footnotesize
\begin{minipage}{\linewidth}
\texttt{Translating Mutable Globals (Strategy C)}
\begin{lstlisting}[language=C, style=customc]
// csplit.c
long	 nfiles;
const char *prefix;
void cleanup(void)
{
    char fnbuf[PATH_MAX];
    ...
    for (i = 0; i < nfiles; i++) {
        snprintf(fnbuf, ..., prefix, ...);
        unlink(fnbuf);
    }
} ...
\end{lstlisting}
\end{minipage}
\begin{minipage}{\linewidth}
\begin{lstlisting}[language=Rust, style=boxed]
// csplit.rs
static COUNT: AtomicU32 = AtomicU32::new(0);
static PREFIX: OnceCell<String> = OnceCell::new();
fn clean_up_handler() -> io::Result<()> {
    ...
    let count = COUNT.load(Ordering::Acquire);
    let pref = PREFIX.get().unwrap();
    for i in 0..count {
        let filename = format!("{}...", pref, ...);
        let _ = fs::remove_file(...); 
    }
    Ok(())
} ...
\end{lstlisting}
\end{minipage}
    \caption{(Top) A part of the \texttt{csplit} C program with mutable globals. (Bottom) A Rust translation using strategy C: atomics.}
    \label{fig:globalatom}
\end{figure}

\todoadd{\emph{(B) Dynamic References.} Fig. \ref{fig:globaldyn} shows an example where global mutable variables are translated into dynamic references. The top of Fig. \ref{fig:globaldyn} is a piece of the \texttt{csplit} C program. The \code{cleanup} is a function to be called as a signal handler. A user translation (bottom of Fig. \ref{fig:globaldyn}) uses a dynamic reference \code{Arc\<Mutex\<T\>\>} to implement this global variable. Correspondingly, the Rust translation uses \code{lock().unwrap()} to guard the access to the content of the variable. It is worth noting that while signal handlers, in general, need synchronization, it is not necessary in this specific program considering its possible behaviors. }

\todoadd{\emph{(C) Atomics.} Fig. \ref{fig:globalatom} shows a different Rust translation from Strategy (C) of the same piece of code in \texttt{csplit}. This time, the code is leveraging an \code{AtomicU32} type declared as a mutable global variable. A nice property of \code{Atomic} types is that ``immutable'' references to it (from the perspective of the Rust language) allow both read and write access to the variable in safe Rust using specific APIs, similar to the interior mutability pattern. Thus, every function can freely obtain ``immutable'' references and write to \code{Atomic} types without violating borrowing rules. At the same time, the runtime cost is also potentially lower than that of locks.}

\begin{figure}[th]
    \centering\footnotesize
\begin{minipage}{\linewidth}
\texttt{Translating Unions as Enum}
\begin{lstlisting}[language=C, style=customc]
// shoco.c
union Code {
  uint32_t word;
  char bytes[4];
};
size_t shoco_compress(...) {
  union Code code;
  while // ...
    // ... code lines modifying code.word
    for (unsigned int i=0; i<packs[pack_n].bytes_packed; ++i)
      o[i] = code.bytes[i];
    // ...
} ...
\end{lstlisting}
\end{minipage}
\begin{minipage}{\linewidth}
\begin{lstlisting}[language=Rust, style=boxed]
// shoco.rs
pub fn shoco_compress(...) {
    let mut code: u32;
    while // ...
        // ... code lines modifying code
        for i in 0..packs[pack_n].bytes_packed {
            if cfg!(target_endian = "little") {
                out[out_index + i] = code.to_le_bytes()[i];
            } else {
                out[out_index + i] = code.to_be_bytes()[i];
            }
        } ...
} ...
\end{lstlisting}
\end{minipage}
    \caption{(Top) A part of the \texttt{shoco} C program that uses tagged unions. (Bottom) A Rust translation that translates tagged union into enum.}
    \label{fig:unionpun}
\end{figure}

\begin{figure}[th]
    \centering\footnotesize
\begin{minipage}{\linewidth}
\texttt{Translating Unions as Enum}
\begin{lstlisting}[language=C, style=customc]
// expr.c
struct val {
    enum { integer, string } type;
    union {
        char    *s;
        int64_t  i;  } u;
};
int is_zero_or_null(struct val *vp)
{
    if (vp->type == integer)
        return vp->u.i == 0;
    else
        return *vp->u.s == 0 || 
          (to_integer(vp, NULL) && vp->u.i == 0);
} ...
\end{lstlisting}
\end{minipage}
\begin{minipage}{\linewidth}
\begin{lstlisting}[language=Rust, style=boxed]
// expr.rs
enum Val {
  Integer(i64),
  String(String),
}
impl Val {
  fn is_zero_or_null(&self) -> bool {
    match self {
      Val::Integer(i) => *i == 0,
      Val::String(s) => s.is_empty() || s.parse::<i64>().map_or(false, |i| i == 0),
    }
  } ...
\end{lstlisting}
\end{minipage}
    \caption{(Top) A part of the \texttt{expr} C program that uses tagged unions. (Bottom) A Rust translation that translates tagged union into enum.}
    \label{fig:unionenum}
\end{figure}

\noindent\todoadd{\textbf{Translating Unions.} There are two different use cases of C union observed in our benchmarks, including (A) type punning and (B) variant record. One example of each is shown here.}

\todoadd{\emph{(A) Type Punning.} Fig. \ref{fig:unionpun} shows an example of type punning in the \texttt{shoco} C program. The C union \code{Code} is used to access bytes of an integer (\code{uint32_t}). An example Rust translation (bottom of Fig. \ref{fig:unionpun}) uses \code{to_le_bytes} and \code{to_be_bytes} that convert an integer (\code{u32}) into a byte array (\code{[u8; 4]}). The choice of API is guarded by a compile-time constant \code{cfg!(target_endian = "big")} that is similar to conditional compilation in C (such as \code{\#ifdef}). The type conversion and the if branch in the Rust code have zero cost when optimization is turned on. Moreover, the differences between the \code{to_le_bytes} and \code{to_be_bytes} are also small after optimization. On x64 architecture (little endian), \code{to_be_bytes} will result in code with just one more \code{bswap} instruction.\footnote{Tested with \code{rustc 1.81.0} with option \code{-C opt-level=2}.}}

\todoadd{\emph{(B) Variant Record.} Fig. \ref{fig:unionenum} shows an example of using tagged unions in C. The top of Fig. \ref{fig:unionenum} shows a slice of the \texttt{expr} C program that declares a variant record (\code{val}) and code accessing the field \code{u} which is a union. From the code, we can infer that semantically there is a mapping between the value of tag \code{type} and the valid field in the union \code{u}. Such a mapping is an important type-safety invariant, but it is not explicit in the type system of C. The Rust translation, shown at the bottom of Fig. \ref{fig:unionenum}, implements the same functionality with statically enforced type safety. The code implements the variant record using \code{enum} in Rust, which explicitly associates tag values to valid fields in the record. Code accessing an \code{enum} in Rust is checked statically, preventing type safety violations as well as logical errors such as missing the handling of certain tags.}

\fi

\end{document}